\documentclass[12pt]{article}
\usepackage{hyperref}
\usepackage{xcolor}
\usepackage{amsmath}
\usepackage{amssymb}
\numberwithin{equation}{section}
\renewcommand{\title}[1]{%
    \bigskip%
    \begin{center}%
    \Large\bf #1%
    \end{center}%
    \vskip .2in}

\renewcommand{\author}[1]{%
    {\begin{center}
    #1
    \end{center}}}
\newcommand{\address}[1]{\vspace{-1.7em}\vspace{0pt}
    {\begin{center}
    \it #1
    \end{center}}}

\begin{document}
\title{Non-relativistic fluids on scale covariant Newton-Cartan backgrounds
}

\author
{
Arpita Mitra }

\vspace{0.5em}

\address{S. N. Bose National Centre 
for Basic Sciences, Block J.D., Sector III, Salt Lake, Kolkata -700 106, India }

\address{\tt arpita12t@bose.res.in}

\begin{abstract} 
The non-relativistic covariant framework for fields is extended to investigate fields and fluids on scale covariant curved backgrounds. The scale covariant Newton-Cartan background is constructed using the localization of spacetime symmetries of non-relativistic fields in flat space. Following this, we provide a Weyl covariant formalism which can be used to study scale invariant fluids. By considering ideal fluids as an example, we describe its thermodynamic and hydrodynamic properties and explicitly demonstrate that it satisfies the local second law of thermodynamics. As a further application, we consider the low energy description of Hall fluids. Specifically, we find that the gauge fields for scale transformations lead to corrections of the Wen-Zee and Berry phase terms contained in the effective action.
% The consequences of these additional terms in the context of Hall phenomenology is discussed.
\end{abstract}
\section{Introduction}
The coupling of non-relativistic field theories to curved backgrounds was initially investigated with the motivation of establishing a covariant framework. In recent years, such minimally coupled theories have gained a renewed importance due to major applications in the field of condensed matter physics, such as in the descriptions of the fractional quantum Hall effect, trapped electron gas, and various transport phenomena, to name a few \cite{Son:2005rv}-\cite{Geracie:2015dea}. Diffeomorphism invariance in non-relativistic systems have certain distinct features which set them apart from the usual relativistic case, with notable difficulties in coupling fields to curved spacetime. In the case of Riemannian spacetime, any arbitrary foliation is acceptable. However in the non-relativistic case, there exists a unique coordinate system which is Galilean invariant in the flat limit. As the non-relativistic background comprise of two degenerate, mutually orthogonal metrics, it also does not lead to a straightforward application of the minimal coupling prescription. %This covariant background, well known as the Newton-Cartan geometry in the literature, was first constructed by Elie Cartan \cite{Cartan:1923zea} and helps in appreciating Newtonian gravity as a non-relativistic limit of General Relativity. However, the current requirement is a formulation based on the vielbeins which will be an analog of the Cartan formulation of Einsteins gravity. 
One may think that a suitable algorithm may be obtained
from relativistic theories by contraction. We also note that the requirement of spatial diffeomorphism in any planar system is difficult to obtain by taking a non-relativistic limit of some appropriate relativistic theory. Moreover such non - relativistic limits are not unique \cite{Wu:2014dha, Banerjee:2016bbm}.
 
A way out follows from the well known procedure for directly deriving relativistic matter theories minimally coupled to curved backgrounds, namely through the localization of spacetime symmetries in the flat spacetime field theory \cite{Utiyama:1956sy, Blagojevic:2002du}. One starts with a matter theory invariant under global Poincar\'e transformations, which fails to remain invariant when the parameters of the transformations are made functions of spacetime. Invariance of the theory under local Poincar\'e transformations follows from the introduction of compensating fields through the definition of covariant derivatives. An important aspect of this approach is the correspondence of these new fields with the vierbeins and spin-connection in Riemann-Cartan spacetime. Following the spirit of this procedure, a general prescription to attain non-relativistic diffeomorphism invariance was proposed in our previous work \cite{Banerjee:2014pya, Banerjee:2015tga}, wherein the localization of the Galilean symmetry for field theories was carried out. The Newton-Cartan spacetime was found to be the most general Galilean covariant curved background through this procedure \cite{Banerjee:2014nja,Banerjee:2015rca}. The present work seeks to extend this formalism to further include the anisotropic scale transformations with dynamical exponent $z=2$ \cite{Hagen:1972pd}.

The main motivation of this paper is to construct non-relativistic scale covariant curved backgrounds and investigate their effect on the description of fluids. In this context, we note that many properties of non-relativistic fluids have been investigated holographically via the fluid-gravity correspondence \cite{Rangamani:2008gi} using the well-known light cone reduction formalism. Non-relativistic fluids on the Newton-Cartan background have in addition been discussed in \cite{Duval:1976ht, Geracie:2015xfa, Geracie:2014nka}. %While relativistic systems scale uniformly in space and time, non-relativistic systems are characterized by an anisotropic scaling between the two \cite{Hagen:1972pd}. It is obvious that the minimal coupling of fields on the scale covariant Newton-Cartan background will differ from the usual relativistic prescription in order to maintain diffeomorphism invariance in the presence of two degenerate metrics. 
  %However, scale invariance  were not considered in these works, and is likely to play an anologously subtle role in the covariant formulation of non-relativistic field theories. 
Despite the consideration and construction of certain non-relativistic conformal backgrounds in \cite{Duval:2009vt, Jensen:2014aia, Bergshoeff:2014uea}, the implications of minimally coupling scale invariant fields to them have attracted limited attention. Our interest in scale invariant fluids stems from its known significance in the relativistic case, where properties of the Riemann tensor and the dilatation field are known to affect hydrodynamic equations~\cite{Loganayagam:2008is}. Scaling also plays an important role in determining temperature dependence of transport coefficients in the hydrodynamic description of condensed matter systems with ordinary critical points \cite{Hohenberg:1977ym}. Scale invariance is also central to the flat space effective field theory descriptions of the fractional quantum Hall effect \cite{Fradkin:1991nr} and the Aharanov-Bohm effect \cite{Bergman:1993kq}. 
%Hall fluids in particular are known to be incompressible in nature \cite{Laughlin:1983fy, Zhang:1992eu}. An important feature of incompressible fluids is that the Euler equations are invariant under the scale transformation but not under the special conformal transformations \cite{Fouxon:2008ik}. %Thus non-relativistic conformal incompressible fluids are only scale invariant, and the formulation we present here would be relevant in their description on curved backgrounds.It is also known that incompressible non-relativistic fluids are invariant under scale transformations only and not under the special conformal transformation \cite{Fouxon:2008ik}.
 %Thus, while scale invariance is a symmetry of all conformal fluids, the full conformal group need not be. Motivated by these observations the present work is being undertaken with the goal of investigating the Newton-Cartan background covariant under the anisotropic scale transformations with dynamical exponent $z=2$. While the Riemann tensor is not invariant under the scale transformations of the metric, a scale invariant Weyl tensor can be constructed out of the Riemann and Ricci tensors.  
 
In the case of relativistic conformal fluids, a Weyl covariant formalism can be constructed from the dilatation field. The formalism leads to the dilatation field being expressed in terms of the expansion and acceleration of the fluid, as well as the covariant expression for the entropy current of conformal fluids on curved backgrounds \cite{Loganayagam:2008is}. The thermodynamic description of these fluids, in effect, is related to the dilatation field on curved backgrounds. Inspired by these results and the aforementioned relevance of scale in the hydrodynamic descriptions of systems near criticality, we expect that understanding scale invariance in non-relativistic fluids on curved backgrounds might serve a broader context. The implications of scale invariance on the incompressible Hall fluid is interesting in its own right. The quantized Hall conductivity is the most fundamental transverse response for Hall fluids. The charge current flows perpendicular to the direction of an external electric field and the transport is dissipationless \cite{Tsui:1982yy}. While the Hall conductivity is a key topological property, independent of the microscopic details of the system, it does not specify the Hall fluid completely. A full characterization of the Hall fluid also requires the intrinsic orbital spin and the corresponding Hall viscosity. These quantities generate from the coupling of the Hall fluid to a curved background. The Hall viscosity is the response of the Hall fluid to an external shear deformation of the background surface, under which the Hall fluid develops a momentum density perpendicular to it. As a result, the net energy for the deformation vanishes resulting in a non-dissipative viscosity. At the level of the effective hydrodynamic theory, this coupling of the hydrodynamic gauge fields of the fluid to the spin connection is represented by the Wen-Zee term (whose coefficient involves the orbital spin) \cite{Wen:1992ej}, while the usual contribution to the Hall viscosity arises from the Berry phase term (which contains the density and the temporal piece of the spin connection) \cite{Cho:2014vfl}. In considering scale transformations, we find the scale analogs of the Wen Zee and Berry phase terms, which involve the dilatation field in place of the spin connection. While the Berry phase and Wen Zee terms provide a Hall viscosity due to the antisymmetric nature of the spin connection, the dilatation field leads to a non-trivial expansion of the fluid. Aspects of this response in the context of the Newton-Cartan background will be further elaborated in the paper.

This paper is organized as follows. Section \ref{App1} details the localization of spacetime symmetries to develop the vierbein formulation of the scale covariant Newton-Cartan geometry. In Section \ref{SNC}, we construct the scale covariant Newton-Cartan background beginning with a short review of some basic properties of the torsion-free Newton-Cartan geometry. In this section we will demonstrate that the curved backgound constructed through the localization of Galilean and scale transformations can be identified with the scale covariant Newton-Cartan geometry. %and the Riemann tensor of the background is not invariant under the anisotropic scale transformation, while the constructed Weyl tensor is. 
In Section \ref{flow}, we develop the Weyl covariant formalism suited to non-relativistic scale invariant fluids following a brief description of the properties of ideal fluids on the Newton-Cartan background. In Section \ref{QHENC}, the effect of non-relativistic scale invariance in the effective hydrodynamic description of the Hall fluid is presented. In Section \ref{Conc}, we conclude with a discussion of our results. Appendix \ref{app1} provides a detailed derivation of Schouten tensor of the Newton-Cartan background. 

\section{Localization of Galilean and scale transformations} \label{App1}

%Relativistic matter theories minimally coupled to curved backgrounds can be derived through the localization of spacetime symmetries in the flat spacetime field theory \cite{Utiyama:1956sy, Blagojevic:2002du}. One starts with a matter theory invariant under global Poincare transformations which does not remain invariant when the parameters of the Poincare transformations become functions of spacetime. To render the theory invariant under the local Poincare transformations, compensating fields are introduced in the process by defining covariant derivatives \cite{Blagojevic:2002du}. An important aspect of this approach is the correspondence of these new fields with the vierbeins and spin-connection of the Riemann-Cartan spacetime. 

 %The Newton-Cartan spacetime was found to be the most general Galilean invariant curved background through this procedure \cite{Banerjee:2014nja,Banerjee:2015rca}. 
The minimal coupling of non-relativistic matter theories to curved background had been considered in \cite{Geracie:2015dea,Duval:1976ht,Duval:1983pb}. Other approaches have been put forward to determine the nature of curved non-relativistic backgrounds directly from the consideration of non-relativistic symmetries. One of these involves the derivation of the background geometry with appropriate metric and curvature tensors, by gauging the centrally extended Galilean algebra (Bargmann algebra) \cite{Andringa:2010it}. The conformal extension of this procedure has been carried out in \cite{Bergshoeff:2014uea}. Yet another approach, which is very closely related to the gauging approach mentioned above is the coset construction \cite{Brauner:2014jaa, Jensen:2014aia}. Central to the success of these approaches, as well as our own, are the presence of vierbeins. Through the involvement of vierbeins, the end result is guaranteed to be manifestly covariant and independent of any specific choice of coordinates. However our proposed prescription of localizing the space-time symmetries \cite{Banerjee:2014pya, Banerjee:2015tga} has a specific advantage. Bearing the non-relativistic nature of absolute time, the parameters of temporal transformations depend only on time and not space. This in turn affects which vierbeins do result from the procedure, and serves to elucidate the relation between the vierbeins and the localization of the parameters one begins with. 

In this section the procedure described in \cite{Banerjee:2014pya, Banerjee:2015rca} will be extended to include anisotropic scale transformations for $z=2$ field theories. The procedure applies to any action invariant under the (global) Galilean and scale transformations in flat space. For general $z$, a prescription has been provided in \cite{Hartong:2015zia}. Let us consider the global action

\begin{equation}
S = \int dt d^3x~ {\cal{L}}\left(\phi, \partial_t \phi, \partial_k \phi\right)\label{genaction}
\end{equation}
where  the index `$t$' and `$k = 1,2,3$' denote time and spatial coordinates respectively. In covariant notation these can be represented collectively by $\mu$.  

The Galilean transformation on time is a translation, while on spatial coordinates it is a composition of spatial translations, rotations and boosts. Its infinitesimal version is as follows,
\begin{equation}
t\rightarrow t-\epsilon,~~~~~~x^i\rightarrow x^{i}+\epsilon^{i}+ \omega^{i}{}_{j}x^{j}-v^{i}t =x^i+\eta^i-v^i t\label{globalgalilean}
\end{equation}
where $\eta^i=\epsilon^{i}+ \omega^{i}{}_{j}x^{j}$, and $\omega^{ij}$ is antisymmetric. The parameters $\epsilon$, $\epsilon^{i}$, $\omega^{ij}$ and $v^{i}$ correspond to time translation, space translation, spatial rotation and boost respectively. For global transformations these parameters are constant.

In non-relativistic systems, time gets rescaled `z' times as compared to the space coordinates \cite{Hagen:1972pd}, where `z' is called the dynamical critical exponent. This is well known as `Lifshitz scaling' in the literature.\footnote{Another non-relativistic conformal extension known in the literature is that of the Galilean Conformal algebra \cite{Bagchi:2009my}. The generator for non-relativistic scaling in GCA is,
\begin{equation}
D =-(x^i\partial_i+t\partial_t)\notag\\
\end{equation}
%K (&=K_0=-(2tx^i\partial_i+t^2\partial_t),K_i &=t\partial_i)
In GCA, space and time scale uniformly like the relativistic case and the number of generators are the same as those of the relativistic conformal group. } It plays an important role in strongly coupled systems, which have been investigated holographically and also found to be relevant in the description of strange metals \cite{Hartnoll:2009ns}. The expression of the scale transformations in time and space coordinates are given by,
\begin{equation}
t'=e^{zs} t,~~~x^{i'}=e^s x^i\label{s}
\end{equation}
where `s' is the parameter of the scale transformation. The infinitesimal transformation takes the following form,
\begin{equation}
x^i\rightarrow x^i+sx^i,~~~t\rightarrow t+zst
\label{gst}
\end{equation}
As a specific case, we will consider the complex Schr\"odinger field theory, whose action is given by,
 \begin{equation}
S = \int dt  \int d^3x  \left[ \frac{i}{2}\left( \phi^{*}\partial_{t}\phi-\phi\partial_t\phi^{*}\right) -\frac{1}{2m}\partial_k\phi^{*}\partial_k\phi\right]
\label{globalaction} 
\end{equation}
This theory is invariant under Eq.~(\ref{gst}) in $(3+1)$d when $z=2$ and provided $\phi$ varies as
\begin{equation*}
\phi'(x')=e^{-\frac{3}{2}s}\phi(x)
\end{equation*}
Thus $\Delta{\cal{L}}$ ($=\delta_0{\cal{L}}+\xi^{\mu}\partial_{\mu}{\cal{L}}+\partial_{\mu}\xi^{\mu}{\cal{L}}$) for (\ref{globalaction}) vanishes under the following variations of the field $\phi$ and its derivatives,
\begin{align}
\delta_0\phi &= -\xi^0\partial_t\phi-\xi^i\partial_i\phi- imv^{i}x_i \phi-\frac{3}{2}s\phi\notag\\
\delta_0\partial_k\phi&=-\xi^0\partial_{t}(\partial_{k}\phi)-\xi^i \partial_{i}(\partial_{k}\phi)-imv^i\partial_{k}(x_i\phi)+\omega_k{}^{m}\partial_{m}\phi-\frac{5}{2}s\partial_k\phi
\notag\\
\delta_0 \partial_t\phi&=-\xi^0\partial_{t}(\partial_{t}\phi)-\xi^i \partial_{i}(\partial_{t}\phi)-imv^i x_i\partial_{t}\phi+v^{i}\partial_{i}\phi-\frac{7}{2}s\partial_t\phi
\label{delphi}
\end{align}
where $\xi^0=-\epsilon+2st$ and $\xi^i=\epsilon^{i}+ \omega^{i}{}_{j}x^{j}-v^{i}t+sx^i$.

Now we will follow a similar localization procedure as that described in \cite{Banerjee:2014pya}. To retain the invariance of the theory under local transformations we first introduce additional fields through the `gauge covariant derivatives' \cite{Banerjee:2014pya}. The covariant derivatives are defined as, 
\begin{eqnarray}
D_k\phi=\partial_k\phi+i\alpha {\omega}_k\phi+i\beta C_k\phi\nonumber\\
D_t\phi=\partial_t\phi+i\alpha {\omega}_t\phi +i\beta C_t \phi\label{firstcov}
\end{eqnarray}
The `${\omega}_{\mu}$' fields account for the change due to the localization of the Galilean transformations \cite{Banerjee:2014pya}, while the `C' fields in (\ref{firstcov}) do so for scale transformations. The coefficients of $\omega$ and $C$ fields are $\alpha$ and $\beta$ respectively, which depend on the matter theory. The explicit structure of the gauge fields `$\omega$' and `$C$' are as follows,
\begin{align}
 \omega_k &= {\omega}_k^{ab}\lambda_{ab} + {\omega}_k^{a0}\lambda_{a}\notag\\
 \omega_t &= {\omega}_t^{ab}\lambda_{ab} + {\omega}_t^{a0}\lambda_{a}\notag\\C_{\mu}&=Db_{\mu}
\label{gaugefields}
\end{align}
where $\lambda_{ab}$ and $\lambda_{a}$ are respectively the generators of rotations and Galilean boosts, and `$D$' is the dilatation generator.  The expression for the generator of the Galilean boost is given by $\lambda_a = mx_a$. In (\ref{gaugefields}) ${\omega}_{\mu}^{ab}$ and ${\omega}_{\mu}^{a0}$ are the spin connections corresponding to rotations and boosts respectively, while $b_{\mu}$ is the dilatation field.

It can be noted that these definitions are insufficient to restore the invariance of the theory under local transformations since $D_k\phi \; \text{and} \; D_t\phi$ do not vary as the partial derivatives in (\ref{delphi}). In order to remedy this, we proceed by introducing local spatial coordinates `$x^a$' (a =1, 2, 3), which will also help in parametrizing the tangent space of the curved background. We can then define the covariant derivatives with respect to these local coordinates in the following way
\begin{align}
\tilde{D}_{0}\phi &={e_0}^{0}D_t \phi+{e_0}^{k}D_k \phi\notag\\
\tilde{D}_a\phi &={e_a}^{k} D_k\phi
\label{finalcov}
\end{align}
where the $e_{\alpha}^{\mu}$ are inverse vierbein fields \cite{Banerjee:2014nja}. Note that local time will be the same as the global one due to the absolute nature of Newtonian time. It can now be demonstrated that the derivative $\tilde{D}_a\phi$ defined in Eq.~(\ref{finalcov}) do transform in a manner similar to the partial derivative $\partial_k\phi$ in (\ref{delphi}), provided the fields $\omega_k, C_k$ and ${e_a}^{k}$ vary according to
\begin{align}
{\delta}_0 \omega_{k}&= -\xi^{\mu}\partial_{\mu}\omega_k-\partial_k\xi^{\mu} \omega_{\mu}+ m{\partial}_kv^ix_i + m(v_k-{e_k}^a v_a)\notag\\
\delta_0 C_k &=-\xi^{\mu}\partial_{\mu}C_k-\partial_{k}\xi^{\mu} C_{\mu}+\frac{1}{2}\partial_k s\notag\\
\delta_0 e_{a}{}^{k}&=-\xi^{\mu}\partial_{\mu}e_{a}{}^{k}+\partial_i\xi^k e_{a}{}^{i}-s e_{a}{}^{k}-\omega^b{}_a e_{b}{}^{k}
\label{delBk}
\end{align}
%where $\xi^0=(\epsilon(t)-\lambda(t) t), \xi^i=\left(\eta^i(x,t)-tv^i(x,t)+s(x,t)x^i\right)$. For invariance of the local action the magnitude of time scaling parameter $\lambda$ will be twice of space scaling parameter $s$.

Likewise, the temporal covariant derivative, $\tilde{D}_0\phi$, would transform as $\partial_t\phi$ provided the variations of $\omega_t, C_t, e_{0}{}^{0}$ and $e_{0}{}^{k}$ satisfy,
\begin{align}
\delta_0 \omega_t &=-\xi^{\mu}\partial_{\mu}\omega_t-\partial_t\xi^{\mu} \omega_{\mu}+me_0^k{v_k}+m\partial_t{v}^i x_i\notag\\
\delta_0 C_t &=-\xi^{\mu}\partial_{\mu}C_t-\partial_{t}\xi^{\mu} C_{\mu}+\frac{3}{2}\partial_t s\notag\\
\delta_0 e_{0}{}^{0}&=-\xi^{\mu}\partial_{\mu} e_{0}{}^{0}+\partial_t\xi^0 e_{0}{}^{0}\notag\\
\delta_0 e_{0}{}^{k} &=-\xi^{\mu}\partial_{\mu}e_{0}{}^{k}+\partial_{\mu}\xi^k e_{0}{}^{\mu}-2s e_{0}{}^{k}+v^b e_b{}^k
\label{delBt}
\end{align}
We can now replace the partial derivatives in the action (\ref{genaction}) with these local covariant derivatives to give,
$$
{{\cal{L}}\left(\phi, \partial_t\phi, \partial_k\phi\right)} \to
 {{\cal{L}^{\prime}}\left(\phi, \tilde{D}_0\phi, \tilde{D}_a\phi\right)}
$$
However, `${\cal{L'}}$' is not invariant under the local Galilean and scale transformations and the invariance of the action requires the change in the measure be accounted for. Therefore the Lagrangian is modified to
\begin{equation}
{\cal{L}}={\Lambda} {\cal{L'}}
\end{equation}
To retain the invariance of the action,
`$\Lambda$' has to satisfy,
\begin{equation}
\delta_0 \Lambda+\xi^{\mu}\partial_{\mu}\Lambda=0
\end{equation}
The appropriate form of `$\Lambda$' is found to be,
\begin{equation}
\Lambda=\frac{1}{e_0{}^0}\text{det} {e_k}^a
\label{mes}
\end{equation}
which is the Jacobian for the Galilean and scale transformations.

Replacing the partial derivatives with the local covariant derivatives and considering the change in the measure, we have the following action, 
\begin{equation}
S = \int dt d^3x ~\Lambda{\cal{L}}\left(\phi, \tilde{D}_0\phi, \tilde{D}_a\phi\right)
\label{localactionold}
\end{equation}
Following the aforementioned procedure for the Schr\"odinger action (\ref{globalaction}), one can achieve the localized form as follows 
 \begin{equation}
S = \int dt  \int d^3x~ \Lambda \left[ \frac{i}{2}\left( \phi^{*}\tilde{D}_0\phi-\phi \tilde{D}_0\phi^{*}\right) -\frac{1}{2m}\tilde{D}_a\phi^{*}\tilde{D}_a\phi\right].
\label{localscaleschrodinger} 
\end{equation}
Note that unlike relativistic theories, the \emph{mass} is not the coefficient of the linear term in the potential, but enters as a passive parameter in the kinetic term which has {\textit{no scaling dimension}} \cite{Bergman}. Since non-relativistic theories hold in the regime where the energies being dealt with are far less than the (rest) mass, massive scale invariant non-relativistic theories can, and do exist. In the following section we will describe how one can identify the curved background derived above as the scale covariant Newton-Cartan geometry.

\section{Construction of the Scale covariant Newton-Cartan geometry} \label{SNC}
One major application of the localization procedure in \cite{Banerjee:2014pya} was the construction of Newton-Cartan geometry through a specific identification of the fields, which was discussed in detail in \cite{Banerjee:2014nja}. %A four dimensional manifold was defined with two coordinate systems, local and global, such that at every global coordinate point there is a local coordinate system. The previously introduced field, $\Sigma_{\alpha}{}^{\mu}$, can be interpreted as the vierbein which maps the global and local frames. It was demonstrated in \cite{Banerjee:2014nja} that the 4-d manifold endowed with $\Sigma_{\alpha}{}^{\mu}$ and its inverse $\Lambda_{\mu}{}^{\alpha}$ had the features of the Newton-Cartan geometry.
As we have seen, the additional inclusion of scale invariance has led to a different result following localization. First, the transformation properties of the additional fields that were introduced at the time of localization of the Galilean symmetry were modified. Second, the localization procedure brought in additional gauge fields that were required in order to render the action invariant. %The gauge fields reduce to those found in the localization of Galilean symmetry when the scale parameters $s, \lambda \to 0$.
 We thus expect on account of the different fields introduced in the localization procedure, each with their own scaling dimension, to lead to a different geometry upon identifying the vierbeins of the manifold. However, this geometric structure should reduce to the Newton-Cartan geometry in the limit of vanishing scale parameters. We will begin with the review of some basic properties of the torsion-free Newton-Cartan background.

\emph{Newton-Cartan geometry}: \label{NC}
 The Newton-Cartan background is Cartan's spacetime formulation of the classical Newtonian theory of gravity \cite{Cartan:1923zea}. It is a non-relativistic manifold which contains a degenerate inverse spatial metric and a degenerate temporal 1-form satisfying the following relations,
\begin{align}
\nabla_{\mu}h^{\alpha \beta} = 0 \qquad & \qquad \nabla_{\mu}\tau_{\nu} = 0 \notag \\
h^{\mu \nu} \tau_{\mu} &= 0
\label{ncmet}
\end{align}
where `$\nabla_{\mu}$' is the covariant derivative associated with a connection $\Gamma$ of the manifold. Therefore $h^{\mu\nu}$ and $\tau_{\mu}$ satisfy metricity conditions seperately and are orthogonal to each other. Given that the metrics are degenerate, their inverses do not exist. Formally, we can define a generalized inverse for the temporal 1-form $\tau^{\mu}$, such that
\begin{equation}
\tau^{\mu}\tau_{\mu} = 1
\end{equation}
There exists a class of $\tau^{\mu}$ which satisfy the above relation, with respect to which we can further define a spatial metric $h_{\mu \nu}$ that satisfies the following relations
\begin{align}
h_{\mu \nu}\tau^{\mu} &= 0 \notag \\
\delta^{\mu}_{\nu} &= h^{\mu \lambda}h_{\lambda \nu} + \tau^{\mu}\tau_{\nu}
\label{ncproj}
\end{align}
where $h^{\mu \lambda}h_{\lambda \nu} = P^{\mu}_{\nu}$ is the projection operator of the Newton-Cartan background. From Eq.~(\ref{ncproj}) we can obtain the variation of $h_{\mu\nu}$ as,
\begin{equation}
\delta h_{\mu\nu}=-2h_{\rho(\mu}\tau_{\nu)}\delta \tau^{\rho}
\end{equation}
%In a similar manner the covariant derivative on $h_{\mu\nu}$ will act in the following way,
%\begin{equation}
%\nabla_{\gamma} h_{\mu\nu}=-2h_{\rho(\mu}\tau_{\nu)}\nabla_{\gamma} \tau^{\rho}
%\end{equation}
Thus variations of $h_{\mu \nu}$ are not independent of $\tau^{\mu}$ and only one of these fields must be considered as independent. We adopt the conventional choice of $\tau^{\mu}$ being the independent field under variations.
 
As the covariant derivative in Eq.~(\ref{ncmet}) is metric compatible with both the metrics, the resultant connection is not uniquely determined by these metrics alone. This allows the Newton-Cartan background to geometrically capture the presence of external forces \cite{D}. The Newton-Cartan geometry can be described in terms of vierbein fields ($e^{0}_{\mu}$, $e^{a}_{\mu}$) and their inverses ($e_{0}^{\mu}$, $e_{a}^{\mu}$), which allows to derive the expression of the connection using the `vierbein postulate',
\begin{align}
\nabla_\mu{e_\nu}^{\alpha} = \partial_{\mu}{e_\nu}^{\alpha} - \Gamma_{\nu\mu}^{\rho}{e_\rho}^{\alpha}
+{\omega}^{\alpha}_{\mu\beta}{e_\nu}^{\beta}=0
 \label{VPNC}
\end{align}
where ${\omega}_{\mu}{}^{\alpha\beta}$ are the spin connections. In contrast with the Poincar\'e case, here ${\omega}_{\mu}{}^{\alpha\beta}$ splits into two distinct parts (${\omega}_{\mu}{}^{ab}, {\omega}_{\mu}{}^{a0}$) \cite{Bergshoeff:2014uea}. With all these considerations, one can derive a linear symmetric connection \cite{Banerjee:2014nja} as, 
\begin{align}
{\Gamma^\rho}_{\nu\mu} &= \tau^{\rho}\partial_{(\mu}\tau_{\nu)} +
\frac{1}{2}h^{\rho\sigma} \Bigl(\partial_{\mu}h_{\sigma\nu}+\partial_{\nu}h_{\sigma\mu} - \partial_{\sigma}h_{\mu\nu}\Bigr)+ h^{\rho\lambda}\tau_{(\mu}K_{\nu) \lambda}\notag \\
&= {\Gamma}'^{\rho}_{\nu\mu} +  h^{\rho\lambda}\tau_{(\mu}K_{\nu) \lambda}
\label{nccon}
\end{align}
${\Gamma}'^{\rho}_{\nu\mu}$ in Eq.~(\ref{nccon}) represents the inertial part of the connection, while the full connection ${\Gamma^\rho}_{\nu\mu}$ contains additional non-inertial forces through the term $K_{\lambda \mu}$ \cite{Duval:1983pb}. %The inclusion of torsion would lead (\ref{nccon}) to involve additional terms. 
Further, while $\Gamma'^{\rho}_{\nu \mu}$ takes the same form for all Newton-Cartan connections, there exist many possible parametrizations of $K_{\lambda \mu}$ which are compatible with the symmetries of the background \cite{Hartong:2015zia}. % for a detailed discussion on possible choices for the general Newton-Cartan connection. 
 The requirement that the symmetric Newton-Cartan connection be the non-relativistic limit of a Riemannian connection uniquely restricts $K_{\lambda \mu}$ to only one of the possible cases. 

Given the (torsion-free) symmetric connection (\ref{nccon}), one can construct the Riemann tensor in the usual way,  
\begin{equation}
[\nabla_{\mu}, \nabla_{\nu}]V^{\lambda}=R^{\lambda}_{\phantom{\lambda}\sigma\mu\nu}V^{\sigma}\label{R}
\end{equation}
For a symmetric Newton-Cartan connection, the Riemann tensor satisfies the following relations,
\begin{equation}
\tau_{\rho}R^{\rho}_{\phantom{\rho}\sigma\mu\nu}=0,~~R^{\lambda}_{\phantom{\lambda}\sigma(\mu\nu)}=0,~~R^{\lambda}_{\phantom{\lambda}[\sigma\mu\nu]}=0,
~~R^{(\lambda\sigma)}{}{}_{\mu\nu}=0
\label{ncRSsymm}
\end{equation}
 If in addition the connection has to possess the correct Newtonian limit of the connection of a Riemannian manifold, then the following additional constraint known as \emph{Trautman's condition} has to be imposed on the Riemann tensor,
\begin{equation}
R^{\lambda}{}_{\sigma}{}^{\mu}{}_{\nu}=R^{\mu}{}_{\nu}{}^{\lambda}{}_{\sigma} 
\label{trautman}
\end{equation}
where all indices in Eq.~(\ref{trautman}) have been raised using $h^{\alpha \beta}$. This condition can be satisfied by requiring that $dK=0$, which implies that
\begin{equation}
K_{\lambda \mu} = 2 \partial_{[\lambda} A_{\mu]}\label{kexp}
\end{equation}
where $A_{\mu}$ is an arbitary 1-form. This form of $K_{\lambda \mu}$ can now be used to provide covariant expressions of Newtonian dynamics. %{\color{purple}{The connection which results from using (\ref{kexp}) corresponds to the general symmetric Newton-Cartan connection being ``linear in $A_{\mu}$" \cite{Hartong:2015zia}.}}
Let us define,
\begin{equation}
\phi = \tau^{\mu}A_{\mu} \, , \qquad h^{\mu \nu}\nabla_{\mu}\nabla_{\nu}\phi = 4 \pi \rho
\label{Poisson}
\end{equation}
Then using Eq.~(\ref{ncRSsymm}) and Eq.~(\ref{kexp}) the Ricci tensor is seen to satisfy the following equation,
\begin{equation}
R_{\mu\nu}=4\pi\rho\tau_{\mu}\tau_{\nu}\label{ncRicci}
\end{equation}
which is the correct Newtonian limit of Einstein's equations. In Eq.~(\ref{ncRicci}) `$\rho$' is the mass density which occurs in Poisson's equation. To further elaborate on how the two form $K$ is associated with forces, we first let the temporal vector $\tau^{\mu}$ be free of accelerations in the inertial frame, i.e. $\tau^{\mu}\nabla'_{\mu}\tau^{\nu} = 0$. Then we find that
\begin{equation}
\tau^{\mu}\nabla_{\mu}\tau^{\nu} = \bar{a}^{\nu} = \tau^{\mu}K_{\mu \lambda}h^{\lambda \nu}
\label{accback}
\end{equation}
Thus $\bar{a}^{\mu}$ is always spatial ($\bar{a}^{\nu}\tau_{\nu} = 0$). Using Eq.~(\ref{accback}) it is now straightforward to demonstrate that
\begin{equation}
-2 h_{\alpha [\mu} \nabla_{\nu]}\tau^{\alpha} = K_{\mu \nu}\label{kkara}
\end{equation} 
%Thus $K_{\mu \nu}$ provides non-inertial forces observed on the Newton-Cartan background.  %The restriction given in Eq.~(\ref{trautman}) will be of importance in our consideration of fluid dynamics later on.  

The Riemann tensor can be used to construct the Weyl tensor, which will be particularly relevant in the context of the scale covariant Newton-Cartan background. The trace free part of the Riemann tensor, known as Weyl tensor, can be constructed as,
\begin{equation}
C_{\lambda\sigma\mu\nu}=R_{\lambda\sigma\mu\nu} +2(h_{\lambda[\mu}S_{\nu]\sigma}+\tau_{\lambda}\tau_{[\mu}S_{\nu]\sigma})-2(h_{\sigma[\mu}S_{\nu]\lambda}
+\tau_{\sigma}\tau_{[\mu}S_{\nu]\lambda})
\label{ncweyl}
\end{equation}
where $`S_{\nu\sigma}$' is the Schouten tensor. However the result does not follow as straightforwardly for the Newton-Cartan background, and the steps involved in determining $S_{\nu \sigma}$ from Eq.~(\ref{ncweyl})has been elaborated on in the Appendix.
%these non-vanishing contractions of the Riemann tensor, and the derivation of the expression of $S_{\nu \sigma}$ are discussed. Therefore both $h_{\sigma\mu}$ and $\tau_{\sigma} \tau_{\mu}$ appear in (\ref{ncweyl}).By requiring that $C_{\lambda\sigma\mu\nu}$ be trace free,  can be derived.The Weyl tensor constructed above finds its utility when scale transformations of the Newton-Cartan fields are considered. In this case the connection of the scale covariant background involves the dilatation field. As a consequence, while the Riemann tensor fails to transform homogeneously under anisotropic scale transformations, the Weyl tensor does. %Thus the Weyl tensor constructed above will be useful in the context of the scale covariant background and scale invariant fields defined on them. In the following section we will consider anisotropic scale transformations and the resulting scale covariant Newton-Cartan background.and demonstrate the effect of dilatations on these relations. We will specifically be concerned with the construction of the Weyl tensor on such backgrounds and its properties in the absence of torsion. This is due to the known importance of the Weyl tensor in the description of general conformal fluids on relativistic backgrounds, which we anticipate will also be the case for nonrelativistic fluids. 
The Newton-Cartan structure described above is invariant under Galilean transformations. We will now consider anisotropic scale transformations with dynamical exponent $z=2$ in addition to the Galilean transformations. The structure which covariantly rescales under this anisotropic scaling will be addressed as the scale covariant Newton-Cartan background.
 
\emph{Scale covariant Newton-Cartan geometry}: We now proceed to discuss the realization of the scale covariant Newton-Cartan geometry from our method of localization of the Galilean and scale symmetry following the same approach described in \cite{Banerjee:2014nja}. The vierbeins and their inverse introduced during localization satisfy,
\begin{align}
e_0^{\mu}e_{\mu}^0 &=1,~~e_{a}^{\mu}e^{a}_{\nu}=\delta^{\mu}_{\nu}+e_0^{\mu}e_{\nu}^0,\notag\\~~ e^{a}_{\mu}e_{b}^{\mu}&=\delta^{a}_{b},~e_0^{\mu}e_{\mu}^a=0,~e^0_{\mu}e_a^{\mu}=0
\end{align}
where $\delta$ refers to the usual Kronecker delta and $\mu,\nu,\cdots$ and $a, b, \cdots$ refer to the curved background and the local tangent space indices respectively \cite{Bergshoeff:2014uea}. %The vierbein formulation will be essential in our treatment of the scale covariant Newton-Cartan background in the following. 
The spatial inverse metric and temporal 1-form of the Newton-Cartan geometry can be written in terms of the vierbeins and their inverses as follows,
\begin{align}
h^{\mu\nu}=e_a^{\mu}e_b^{\nu}\delta^{ab},~~\tau_{\mu}=e_{\mu}^0\label{ncvib}
\end{align}
Following this the spatial covariant metric and the inverse of the temporal 1-form will be,
\begin{equation}
h_{\mu\nu}=e_{\mu}^ae_{\nu}^b\delta_{ab},~\tau^{\mu}=e_0^{\mu}
\end{equation}
We can now rewrite Eq. (\ref{localscaleschrodinger}) as the following covariant action using the definition (\ref{finalcov}),
\begin{equation}
S = \int dt  \int d^3x \sqrt{h} \left[ \frac{i}{2}\left( \phi^{*}\tau^{\mu}{D}_{\mu}\phi-\phi \tau^{\mu}{D}_{\mu}\phi^{*}\right) -\frac{1}{2m}h^{\mu\nu}D_{\mu}\phi^{*}D_{\nu}\phi\right]
\label{diffschrodinger2} 
\end{equation}
%Note that for scalar field `$\tilde{\nabla}_{\mu}$' can be identified with the covariant derivative `$D_{\mu}$' defined earlier \cite{Banerjee:2014nja}. 
The action (\ref{diffschrodinger2}) can now be interpreted as that of a massive Schr\"odinger complex scalar field coupled to the scale covariant Newton-Cartan geometry. 

Due to inclusion of this scale transformation, the Newton-Cartan contravariant spatial metric and covariant temporal 1-form do not obey the metricity conditions. These relations follow from the vierbein postulate satisfied by the vierbeins, %introduced during the localization of symmetries of flat space field theories,
\begin{align}
\tilde{\nabla}_\mu{e_\nu}^{0} &= \partial_{\mu}{e_\nu}^{0} - {\tilde{\Gamma}}_{\nu\mu}^{\rho}{e_\rho}^{0}
+{\omega}^{0}_{\mu\beta}{e_\nu}^{\beta}+2b_{\mu}{e_\nu}^{0} =0\notag\\ 
{\tilde{\nabla}}_\mu{e_\nu}^{a} &= \partial_{\mu}{e_\nu}^{a} - {\tilde{\Gamma}}_{\nu\mu}^{\rho}{e_\rho}^{a}
+{\omega}^{a}_{\mu\beta}{e_\nu}^{\beta}+b_{\mu}{e_\nu}^{a} =0\ 
 \label{P}
\end{align}
where $\tilde{\nabla}$, $\tilde{\Gamma}$ are the covariant derivative and the connection of the scale covariant Newton-Cartan geometry respectively, ${\omega}_{\mu}{}^{\alpha\beta}$ ($\alpha=0,a$) are the same as those of the Newton-Cartan background (\ref{VPNC}) and $b_{\mu}$ is the dilatation field introduced earlier in (\ref{gaugefields}). %Compared to Poincar\'e case here ${\omega}_{\mu}{}^{\alpha\beta}$ splits into spatial and temporal part (${\omega}_{\mu}{}^{ab}, {\omega}_{\mu}{}^{a0}$) \cite{Andringa:2010it}. 
The contravariant spatial metric and covariant temporal 1-form can be expressed in terms of the vierbeins following Eq.~(\ref{ncvib}). Similar to the Newton-Cartan case, Eq.~(\ref{P}) directly leads to the expression for the covariant derivative on $\tau_{\nu}$ and $h^{\mu\nu}$ respectively, 
\begin{equation}
\tilde{\nabla}_{\mu} \tau_{\nu} = - 2b_{\mu}\tau_{\nu}, \tilde{\nabla}_\mu h^{\rho\sigma}=2b_{\mu}h^{\rho\sigma}
\label{tnmetricity}
\end{equation}
Despite the non-metricity, the following orthogonality and projection relations are still satisfied by the scale covariant Newton-Cartan background,
\begin{align}
h^{\mu\nu}\tau_\nu =0,~h_{\mu\nu}\tau^\nu=0,~h^{\mu\lambda}h_{\lambda\nu} = \delta^\mu_\nu - \tau^\mu\tau_\nu,~\tau^{\mu}\tau_{\mu}=1
\end{align}
%In the context of the covariant derivative, the explicit form of the connection can be determined.
 Following the vierbein postulate (\ref{P}) the general expression for the connection can be obtained as,
\begin{equation}
\tilde\Gamma_{\nu\mu}^{\rho} = \partial_{\mu}{e_{\nu}}^\alpha {e_\alpha}^{\rho}
+{\omega}^{\alpha}{}_{\mu\beta}{e_{\nu}}^\beta{e_\alpha}^{\rho}+2b_{\mu}{e_{\nu}}^{0} {e_0}^{\rho}+b_{\mu}{e_{\nu}}^{a} {e_a}^{\rho}\label{vpcon}
\end{equation}
From the first relation of Eq.~(\ref{P}), we can also find the following relation,
\begin{equation}
\partial_{[\mu}\tau_{\nu]} =
\frac{\tilde{T}^{\rho}_{\nu\mu}}{2}\tau_{\rho} -2b_{[\mu}\tau_{\nu]}\label{wtor}
\end{equation} 
where ${\tilde{T}^\rho}_{\nu\mu} = 2\tilde{\Gamma}^\rho_{[\nu\mu]}$ is the torsion tensor of the scale covariant Newton-Cartan background. For the Newton-Cartan background, the temporal component of the torsion tensor $\left( T^{\rho}_{\mu\nu}\tau_{\rho} \right)$ vanishes if $d\tau = 0$ (since $\partial_{[\mu}\tau_{\nu]} =
\frac{T^{\rho}_{\nu\mu}}{2}\tau_{\rho}$) \cite{Banerjee:2016laq}. In including scale transformations the temporal component of the torsion tensor involves additional constraints. Due to the presence of the $b_{\mu}$ field, $\tilde{T}^{\rho}_{\mu \nu} \tau_{\rho} \neq 0$ even while $d\tau = 0$. The second implication is that when $\tilde{T}^{\rho}_{\mu \nu} \tau_{\rho} = 0$ we have the following condition on $d\tau$,
\begin{equation}
\partial_{[\mu}\tau_{\nu]} =- 2b_{[\mu}\tau_{\nu]}
\label{con1}
\end{equation}
This equation implies that for the scale covariant Newton-Cartan background $\tau_{\mu}$ is not closed and the temporal part of the field $b_{\mu}$ can be gauged away. However, Eq.~(\ref{con1}) still leads to the Frobenius condition being satisfied, ensuring the existence of spatial hypersurfaces orthogonal to $\tau_{\mu}$. This relation leads to the ``twistlessness condition" of the TTNC background \cite{Bergshoeff:2014uea}.  

The above cases dealt with the temporal component of the torsion tensor. The general expression of the torsion tensor for the scale covariant Newton-Cartan background, including the spatial component, can be derived from Eq.~(\ref{P}) through
\begin{align}
\partial_{[\mu}e_{\nu]}{}^{\alpha}-\tilde{\Gamma}^{\rho}_{[\nu\mu]}e_{\rho}{}^{\alpha}+{\omega}^{\alpha}_{[\mu\vert\beta\vert}e_{\nu]}{}^{\beta}+2b_{[\mu}e_{\nu]}{}^0\delta_{0}^{\alpha}+b_{[\mu}e_{\nu]}{}^b\delta_{b}^{\alpha}=0
\end{align}
Contracting with $e_{\alpha}{}^{\sigma}$ on both sides results in
%\begin{align}
%\tau^{\sigma}\partial_{[\mu}\tau_{\nu]}+e_{a}{}^{\sigma}(\partial_{[\mu}e_{\nu]}^a+{\omega}^a_{[\mu\vert\beta\vert}e_{\nu]}{}^{\beta})+b_{[\mu}\tau_{\nu]}\tau^{\sigma}+b_{[\mu}\delta_{\nu]}^{\sigma}=
%\frac{\tilde{T}^{\sigma}_{\nu\mu}}{2}\label{antvp}
%\end{align}
%where ${T^\rho}_{\nu\mu} = 2\tilde{\Gamma}^\rho_{[\nu\mu]}$ is known as the torsion tensor.
%Manipulating the terms in the parenthesis one can write the general torsion tensor as
\begin{align}
\frac{\tilde{T}^{\sigma}_{\nu\mu}}{2} &=\left[\tau^{\sigma}\partial_{[\mu}\tau_{\nu]}+2b_{[\mu}\tau_{\nu]}\tau^{\sigma}
\right]\notag\\&\qquad+\left(\partial_{[\mu}e_{\nu]}{}^a+{\omega}^a_{[\mu\vert b}e_{\nu]}{}^b+b_{[\mu}e_{\nu]}{}^a\right)e^{\sigma}_a+K_{\gamma[\nu}\tau_{\mu]}h^{\sigma\gamma}
\label{torfull}
\end{align}
where
\begin{equation}
K_{\gamma[\nu}\tau_{\mu]}h^{\sigma\gamma}=\frac{1}{2} e_{\gamma}^a[\omega^a_{0\nu}\tau_{\mu}
-\omega^a_{0\mu}\tau_{\nu}]h^{\sigma\gamma}
\end{equation}
In Eq.~(\ref{torfull}) the first and second lines represent the temporal and spatial contributions respectively. In the limit of $b_{\mu}\rightarrow 0$, Eq.~(\ref{torfull}) reduces to the general Newton-Cartan torsion. We also note that the torsion-free scale covariant Newton-Cartan background will always be subject to Eq.~(\ref{con1}).

 We can now express the connection in terms of the metrics and the dilatation field ($b_{\mu}$) defined earlier. Making use of (\ref{vpcon}), the symmetric part of connection can be written as,
\begin{align}
\frac{1}{2}[\tilde\Gamma_{\nu\mu}^{\rho}+\tilde\Gamma_{\mu\nu}^{\rho}]& =\frac{1}{2}\left[(\partial_{\mu}{e_{\nu}}^{0} {e_0}^{\rho}+\partial_{\nu}{e_{\mu}}^0 {e_0}^{\rho})+(\partial_{\mu}{e_{\nu}}^a {e_a}^{\rho}+\partial_{\nu}{e_{\mu}}^a {e_a}^{\rho})\right.\notag\\
 &\qquad+({\omega}^{a}_{\mu 0}{e_{\nu}}^{0}{e_a}^{\rho}+{\omega}^{a}_{\nu 0}{e_{\mu}}^0{e_a}^{\rho})+({\omega}^{a}_{\mu b}{e_{\nu}}^b{e_a}^{\rho}+{\omega}^{a}_{\nu b}{e_{\mu}}^b{e_a}^{\rho})\notag\\ & \qquad\left.+2(b_{\mu}{e_{\nu}}^0 {e_0}^{\rho}+b_{\nu}{e_{\mu}}^0 {e_0}^{\rho})+(b_{\mu}{e_{\nu}}^a {e_a}^{\rho}+b_{\nu}{e_{\mu}}^a {e_a}^{\rho})\right]
 \label{PP}
\end{align}
Using ${e_a}^{\rho}=h^{\rho\sigma}{e_{\sigma}}^a$, the above expression will take the following form,
\begin{align}
\frac{1}{2}[\tilde\Gamma_{\nu\mu}^{\rho}+\tilde\Gamma_{\mu\nu}^{\rho}]&=\tau^{\rho}\partial_{(\mu}\tau_{\nu)}+
\frac{1}{2}h^{\rho\sigma}[\partial_{\mu} h_{\sigma\nu}-{e_{\nu}{}^a}\partial_{\mu}{e_{\sigma}}^a]+\frac{1}{2}h^{\rho\sigma}[
\partial_{\nu} h_{\sigma\mu}-{e_{\mu}{}^a}\partial_{\nu}{e_{\sigma}}^a]\notag\\& +\frac{1}{2}({\omega}^{a}_{0\mu}{e_{\nu}}^{0}{e_a}^{\rho}+{\omega}^{a}_{0\nu}{e_{\mu}}^0{e_a}^{\rho}+{\omega}^{a}_{\mu b}{e_{\nu}}^b{e_a}^{\rho}+{\omega}^{a}_{\nu b}{e_{\mu}}^b{e_a}^{\rho})\notag\\&+\frac{1}{2}(b_{\mu}\delta_{\nu}^{\rho}+b_{\nu}\delta_{\mu}^{\rho}+b_{\mu}\tau_{\nu}\tau^{\rho}
+b_{\nu}\tau_{\mu}\tau^{\rho})
\label{comid}
\end{align}
Since we are now considering the symmetric part of $\tilde\Gamma_{\nu\mu}^{\rho}$, we have the following relation,
\begin{align}
-{e_{\nu}}^a\partial_{\mu}{e_{\sigma}}^a
-{e_{\mu}{}^a}\partial_{\nu} e_{\sigma}^a=(-\partial_{\sigma}h_{\mu\nu}-{\omega}^{a}_{\mu b}{e_{\nu}}^b{e_{\sigma}}^a-{\omega}^{a}_{\nu b}{e_{\mu}}^b{e_{\sigma}}^a)+(b_{\mu}h_{\sigma\nu}+b_{\nu}h_{\sigma\mu}-2b_{\sigma}h_{\nu\mu})
\label{conec}
\end{align}
Using this expression in (\ref{comid}) the symmetric part of the connection of the scale covariant Newton-Cartan geometry can be written as 
\begin{align}
\frac{1}{2}[\tilde\Gamma_{\nu\mu}^{\rho}+\tilde\Gamma_{\mu\nu}^{\rho}] & = \tau^{\rho}\partial_{(\mu}\tau_{\nu)} +
\frac{1}{2}h^{\rho\sigma} \Bigl(\partial_{\mu}h_{\sigma\nu}+\partial_{\nu}h_{\sigma\mu} - \partial_{\sigma}h_{\mu\nu}\Bigr)+(b_{\mu}\delta_{\nu}^{\rho}
+b_{\nu}\delta_{\mu}^{\rho}-b_{\sigma}h^{\rho\sigma}h_{\nu\mu})\notag\\& \qquad \qquad + h^{\rho\lambda}\tau_{(\mu}K_{\nu) \lambda}
\label{wrcon}
\end{align}
where the last term is given by \cite{Banerjee:2014nja}
\begin{align}
h^{\rho\lambda}\tau_{(\mu}K_{\nu)\lambda}
 =\frac{1}{2}h^{\rho\lambda}[\tau_{\mu}{\omega}^{a}{}_{0\nu}{e_{\lambda}}^a +\tau_{\nu}{\omega}^{a}{}_{0\mu}{e_{\lambda}}^{a}]
 \label{k}
\end{align} 
%It is evident from Eq.~(\ref{wrcon}) that in the limit of vanishing `b', the expression reduces to that of the symmetric Newton-Cartan connection (\ref{nccon}). 
In the presence of torsion (\ref{torfull}) the general connection becomes, 
 \begin{align}
{\tilde{\Gamma}^\rho}_{\nu\mu} & = \tau^{\rho}\partial_{(\mu}\tau_{\nu)} +
\frac{1}{2}h^{\rho\sigma} \Bigl(\partial_{\mu}h_{\sigma\nu}+\partial_{\nu}h_{\sigma\mu} - \partial_{\sigma}h_{\mu\nu}\Bigr)+(b_{\mu}\delta_{\nu}^{\rho}
+b_{\nu}\delta_{\mu}^{\rho}-b_{\sigma}h^{\rho\sigma}h_{\nu\mu})\notag\\&\qquad+ h^{\rho\lambda}\tau_{(\mu}K_{\nu) \lambda}+\frac{1}{2}h^{\rho\sigma}\left[ -\tilde{T}_{\mu\nu\sigma}-\tilde{T}_{\nu\mu\sigma}+\tilde{T}_{\sigma\nu\mu}\right]
\label{Vtor}
\end{align}
where $\tilde{T}^{\sigma}_{\nu\mu}$ was already defined in Eq.~(\ref{torfull}).

We will now strictly assume that the connection of scale covariant Newton-Cartan background is symmetric ($\tilde{T}^{\sigma}_{\nu\mu}=0$) and will consider the curvature terms and their properties in the following. This construction will be applicable to $z=2$ theories only. For convenience, the symmetric connection will be written in the following way,
\begin{equation}
\tilde{\Gamma}^{\rho}_{\nu\mu} = \Gamma^{\rho}_{\nu\mu}+(b_{\mu}\delta_{\nu}^{\rho}
+b_{\nu}\delta_{\mu}^{\rho}-b_{\sigma}h^{\rho\sigma}h_{\nu\mu})
\label{wrcon2}
\end{equation}
where $\Gamma^{\rho}_{\nu\mu}$ represents the usual Newton-Cartan symmetric connection (\ref{nccon}). The Riemann tensor for the symmetric connection in Eq.~(\ref{wrcon2}) is defined in the usual way,
\begin{equation}
[\tilde{\nabla}_{\mu}, \tilde{\nabla}_{\nu}]V^{\lambda}=\tilde{R}^{\lambda}_{\phantom{\lambda}\sigma\mu\nu}V^{\sigma}\label{WeylR}
\end{equation}
Upon expansion, we find the following result
\begin{align}
\tilde{R}^{\lambda}_{\phantom{\lambda}\sigma\mu\nu}&=R^{\lambda}_{\phantom{\lambda}\sigma\mu\nu}+2\nabla_{[\mu}(b_{\nu]}\delta^{\lambda}_{\sigma}+\delta^{\lambda}_{\nu]}b_{\sigma}-h_{\nu]\sigma}b_{\delta}h^{\delta\lambda})+2\delta^{\lambda}_{[\mu}(b_{\nu]}b_{\sigma}-h_{\nu]\sigma}b_{\rho}b_{\sigma}h^{\rho\sigma})\notag\\&+2b_{\rho}h^{\rho\lambda}b_{[\mu}h_{\nu]\sigma}-
2b_{\rho}\tau^{\rho}\tau_{[\mu}h_{\nu]\sigma}b_{\gamma}h^{\gamma\lambda}
\label{WRiem}
\end{align} 
For the Newton-Cartan background $\tau_{\lambda}R^{\lambda}_{\phantom{\lambda} \sigma \mu \nu} = 0$ allowed us to use $R_{\lambda\sigma\mu\nu} = h_{\lambda \rho} R^{\rho}_{\phantom{\rho}\sigma\mu\nu}$. Here $\tilde{R}^{\lambda}_{\phantom{\lambda}\sigma\mu\nu}$ does not satisfy the properties given in Eq.~(\ref{ncRSsymm}) and (\ref{trautman}). We require $\delta^{\mu}_{\lambda}\tilde{R}^{\lambda}_{\phantom{\lambda}\sigma\mu\nu} = \tilde{R}_{\sigma \nu}$ which implies that one can lower the indices with the combination $(h_{\mu \nu} + \tau_{\mu}\tau_{\nu})$ and raise with $(h^{\mu \nu} + \tau^{\mu} \tau^{\nu})$ using the identity, 
\begin{equation}
\delta^{\mu}_{\lambda} = (h^{\mu \alpha} + \tau^{\mu} \tau^{\alpha})(h_{\alpha \lambda} + \tau_{\alpha}\tau_{\lambda})
\end{equation}
Therefore for the scale covariant Newton-Cartan background, we have
\begin{align}
(h_{\lambda\epsilon}+\tau_{\lambda}\tau_{\epsilon})\tilde{R}^{\lambda}_{\phantom{\lambda}\sigma\mu\nu}&=\tilde{R}
_{\epsilon\sigma\mu\nu}=R_{\epsilon\sigma\mu\nu}+2(h_{\epsilon\sigma}+\tau_{\epsilon}\tau_{\sigma})
\nabla_{[\mu}b_{\nu]}+2(h_{\epsilon[\nu}\nabla_{\mu] }b_{\sigma}+\tau_{\epsilon}\tau_{[\nu}\nabla_{\mu]}
b_{\sigma})\notag\\&\qquad-2\nabla_{[\mu}(h_{\nu]\sigma}b_{\epsilon})+2\tau^{\delta}\tau_{\epsilon}\nabla_
{[\mu}(h_{\nu]\sigma}b_{\delta})+2h_{\epsilon[\mu}b_{\nu]}b_{\sigma}+2\tau_{\epsilon}\tau_{[\mu}b_{\nu]}
b_{\sigma}\notag\\&-2h_{\epsilon[\mu}h_{\nu]\sigma}h^{\gamma\rho}b_{\gamma}b_{\rho}-2\tau_{\epsilon}
\tau_{[\mu}h_{\nu]\sigma}h^{\gamma\rho}b_{\gamma}b_{\rho}+2b_{\epsilon}b_{[\mu}h_{\nu]}\sigma-2b_{\rho}
\tau^{\rho}\tau_{\epsilon}b_{[\mu}h_{\nu]\sigma}\notag\\&-2b_{\rho}\tau^{\rho}\tau_{[\mu}h_{\nu]\sigma}b_{
\epsilon}+2\tau^{\gamma}\tau^{\rho}b_{\gamma}b_{\rho}\tau_{\epsilon}\tau_{[\mu}h_{\nu]\sigma}\label{tilderiemann}
\end{align}
By contracting \ref{tilderiemann} with $(h^{\epsilon \mu} + \tau^{\epsilon} \tau^{\mu})$ we get the expression for the Ricci tensor $\tilde{R}_{\sigma \nu}$, 
\begin{align}
\tilde{R}_{\sigma\nu}&=R_{\sigma\nu}+2\nabla_{[\sigma}b_{\nu]}-\nabla_{\mu}(h_{\nu\sigma}b_{\epsilon}h^{\epsilon\mu})+(n-2)[b_{\nu}b_{\sigma}-\nabla_{\nu}b_{\sigma}-h_{\nu\sigma}h^{\gamma\rho}b_{\gamma}b_{\rho}]
\notag\\&-\tau_{\sigma}\nabla_{\nu}(\tau^{\rho}b_{\rho})+2b_{\rho}\tau^{\rho}\tau_{(\sigma}b_{\nu)}-(b_{\rho}\tau^{\rho})(b_{\gamma}\tau^{\gamma})(\tau_{\nu}\tau_{\sigma})
\label{tildeRicc}
\end{align}
This is of course the same result one would get from Eq.~(\ref{WRiem}) by setting `$\lambda = \mu$'. Again contracting Eq.~(\ref{tildeRicc}) with $(h^{\sigma \nu} + \tau^{\sigma} \tau^{\nu})$ the expression of the Ricci scalar can be obtained as follows,
\begin{align}
\tilde{R}&=R-h^{\mu\nu}\nabla_{\mu}b_{\nu}(2n-3)-(\tau^{\mu}\nabla_{\mu}(b_\rho\tau^{\rho})-
\tau^{\gamma}\tau^{\rho}b_{\gamma}b_{\rho})(n-1)\notag\\&+(n-2)b_{\sigma}\tau^{\rho}\nabla_{\rho}\tau^{\sigma}-
(n-2)^2h^{\gamma\rho}b_{\gamma}b_{\rho}
\label{tildeRiccs}
\end{align}
It is evident from the previous expressions (\ref{tilderiemann}-\ref{tildeRiccs}) that the Riemann tensor, Ricci tensor and Ricci scalar are not invariant under the scale transformations and certain symmetries of the Newton-Cartan Riemann tensor are not satisfied by the rescaled counterpart. If we now assume that %$b_{\mu} = \partial_{\mu} s$ for any scaling parameter $s$ and 
$2 b_{[\mu} \tau_{\nu]} = -\partial_{[\mu} \tau_{\nu]}$ then it can be observed that the Weyl tensor defined in Eq.~(\ref{ncweyl}) is invariant under non-relativistic scale transformations,
\begin{equation}
C^{\lambda}_{\phantom{\lambda}\sigma \mu \nu} = \tilde{C}^{\lambda}_{\phantom{\lambda}\sigma \mu \nu}
\end{equation}

 %These violations are due to the presence of the $b$ fields, which are both spatial and temporal. But other symmetries may be imposed on the Riemann and Ricci tensors described above. For instance, Eq.~(\ref{tildeRicc}) reveals that the Ricci tensor is not symmetric. At this stage we could require the symmetries $\tilde{R}_{[\sigma \nu]} = 0 = \tilde{R}^{\lambda}_{\phantom{\lambda} \lambda \mu\nu}$ to hold for the Riemann tensor of the  rescaled Newton-Cartan background. This in turn determines conditions on the `b' fields through which these symmetries are satisfied. In General Relativity this simply leads to the condition that `$b_{\sigma} = \partial_{\sigma} \alpha$', for some scalar field $\alpha$. Here, apart from this constraint, the additional requirement of `$b_{[\mu} \tau_{\nu]} = 0$' needs to be satisfied. This constraint will be satisfied whenever the spatial hypersurfaces satisfy `{\textit{Frobenius' theorem}}' \cite{Bergshoeff:2014uea}. This point will be further clarified in out consideration of torsion at the end of this section.
However, in many cases it may be useful to consider the symmetries of the rescaled Riemann tensor without imposing additional conditions. For instance, this is useful in the treatment of non-ideal conformal fluids on curved backgrounds \cite{Loganayagam:2008is}. We will briefly discuss this point in the treatment of non-relativistic fluids in the next section. %Note that the above construction and results have followed from the consideration of.\textit{Connection with torsion:} The general Newton-Cartan background can also have torsion, and here we provide a brief account of the implications of scale invariance on it.  %This expression has not been considered in the literature thus far. The spatial contribution
\section{Non-relativistic fluids on curved backgrounds} \label{flow}
%%%%%%%%%%%%%%%%%%%%%%%%%%%%%%%%%%

The aim of this section is to elaborate on an important application of the construction of the previous sections, namely, in the description of non-relativistic fluids. Fields close to equilibrium admit a hydrodynamic description. Within this description the stress tensor and symmetry currents are expressed in a gradient expansion of the fluid variables and the spacetime background.  We will first give a detailed description of ideal fluids on the Newton-Cartan background. Following this we develop a Weyl-covariant formalism which will facilitate the study of scale invariant non-relativistic fluids. %In particular we will consider  fluids on scale covariant backgrounds. 
While our analysis will be confined to the case of ideal fluids, the framework can be used for a more detailed investigation of the derivative expansion involved for general non-relativistic fluids. %In the last section we will investigate some consequences of scale covariant backgrounds on the response functions of Hall fluids.

%The aim of this section is to elaborate on the description of non-relativistic conformal incompressible fluids coupled to a curved background. In the following subsection we will briefly discuss the basic characteristics of fluids on the Newton-Cartan background following \cite{Duval:1976ht, Geracie:2015xfa, Geracie:2014nka}. We will then develop a Weyl-covariant formalism which simplifies the study of conformal non-relativistic hydrodynamics analogous to the relativistic case \cite{Loganayagam:2008is}. This helps to describe the system as a thermodynamical one satisfying the second law of thermodynamics. 
\subsection{Fluid dynamics on the Newton-Cartan background} \label{varis}
%A preliminary study of non-relativistic fluids on the usual was performed in   in the next subsection and then extend them to the conformal case in the following subsection. We will base our subsequent calculations on this work and extend them to the conformal case. 
In the non-relativistic hydrodynamics regime, the basic fluid variables are the local velocity $v^i(x)$ and mass density $\rho(x)$, and they satisfy the following conservation equations,
\begin{align}
\partial_t{\rho}+\partial_i(\rho v^i)&=0~~~ \text{(Continuity equation)}\notag\\
\partial_t(\rho v^i)+\partial_{i}T^{ij}&=0~~~\text{(Momentum conservation equation)}\notag\\
\partial_t\left(\epsilon+\frac{1}{2}\rho {\bf{v}}^2\right)+\partial_i j^i&=0~~~\text{(Energy conservation equation)}
\end{align}
where $T^{ij}, \epsilon$ and $j^i$ are the stress tensor, energy density and matter current of fluid respectively. 

%A preliminary study of non-relativistic fluids on the usual NC background was performed in .
In this subsection we review some of the relevant properties of ideal non-relativistic fluids on the Newton-Cartan background with zero torsion following \cite{Duval:1976ht, Geracie:2015xfa, Geracie:2014nka}. This description requires a choice of fluid velocity, for which we consider $u^{\mu}$ such that 
\begin{equation}
u^{\mu}\tau_{\mu}=1 
\label{vel}
\end{equation} 
A sensible requirement is that the ideal fluid has no acceleration and is irrotational \cite{Duval:1976ht} when considered with respect to the inertial frame of the Newton-Cartan background, i.e.
\begin{equation}
a'^{\mu}=u^{\rho}\nabla'_{\rho}u^{\mu}=0,~~~\omega'^{\mu\nu}=h^{\gamma[\mu}\nabla'_{\gamma}u^{\nu]}=0
\label{eq.inertial}
\end{equation}
where $\nabla'$ is the covariant derivative corresponding to the inertial piece of the Newton-Cartan connection $\Gamma'$ in Eq.~(\ref{nccon}). The total covariant derivative will act on the fluid velocity $u^{\nu}$ as,
\begin{equation}
\nabla_{\mu} u^{\nu} =\nabla'_{\mu} u^{\nu}+h^{\nu\lambda}\tau_{(\mu}K_{\rho) \lambda}u^{\rho}
\label{eq.Kun}
\end{equation}
From Eq.~(\ref{eq.inertial}) and Eq.~(\ref{eq.Kun}) it then follows that the fluid variables for the expansion, acceleration, shear and vorticity for a general Newton-Cartan background can be written as,
\begin{align}
\theta &= \nabla_{\mu} u^{\mu} = \nabla'_{\mu} u^{\mu} = \theta' \notag\\
a^{\nu} &= u^{\mu}\nabla_{\mu} u^{\nu} = h^{\nu \lambda} K_{\rho \lambda} u^{\rho}\notag\\ \sigma^{\mu\nu}&= [h^{ \lambda(\mu}\nabla_{\lambda}u^{\nu)}]-\frac{\theta}{n-1}h^{\mu\nu} = [h^{ \lambda(\mu}\nabla'_{\lambda}u^{\nu)}]-\frac{\theta}{n-1}h^{\mu\nu} = \sigma'^{\mu \nu}\notag \\
\omega^{\mu\nu} &= [h^{ \lambda[\mu} \nabla_{\lambda}u^{\nu]}]= \omega'^{\mu \nu}=0 \label{fluvar}
\end{align}
Thus apart from the acceleration, all other quantities to describe the fluid are invariant in going from an inertial to a non-inertial frame. 
In addition to these quantities, the description of a fluid requires a definition of the stress-energy tensor and other matter currents of the theory. Since the Newton-Cartan background contains two degenerate metrics ($h^{\mu\nu}, \tau_{\mu}\tau_{\nu}$) and additional gauge fields ($h_{\mu\nu}, \tau^{\mu}, A_{\mu}$), these definitions should follow from a careful variation of the action. The gauge field $A_{\mu}$ was introduced in Eq.~(\ref{kexp}).
%To reformulate the fundamental equations of fluid mechanics in a Weyl covariant form we have to discuss the basic equations like conservation of the energy momentum tensor and the matter flow current.
The most general variation of a matter action on the Newton-Cartan background which leaves the connection invariant is given by
\begin{align}
0=\delta S=\int\sqrt{h}d^4x  \left[-\frac{1}{2}P_{\mu\nu}\delta h^{\mu\nu}+Q^{\mu}\delta \tau_{\mu}+J^{\mu}\delta A_{\mu}+R_{\mu}\delta \tau^{\mu}\right]\label{genvar}
\end{align}
where $P_{\mu \nu}, Q^{\mu}, J^{\mu}$ and $R_{\mu}$ will be identified with the physical stress tensor, energy current, mass conservation current and momentum current respectively. Two of these variations correspond to non-gauge variables, i.e. $\delta h^{\mu \nu}$ and $\delta \tau_{\mu}$, which are the variations of the given inverse spatial metric and temporal 1-form. Setting these variations to vanish provides the contributions from the pure gauge variables $A_{\mu}$ and $\tau^{\mu}$. Eq.~(\ref{genvar}) then reduces to,
\begin{equation}
\delta S=\int\sqrt{h}d^4x [J^{\mu}\delta A_{\mu}+R_{\mu}\delta \tau^{\mu}]\label{gac2}
\end{equation}
We can simplify Eq.~(\ref{gac2}) further by using Eq.~(\ref{kkara}) to find the following properties of $K_{\lambda\mu}$,
\begin{equation}
\delta K_{\lambda\mu}= 2\nabla_{[\lambda}h_{\mu]\nu}\delta \tau^{\nu}\label{Kexpr2}
\end{equation}
Since we are working up to first order in variations, it follows from Eq.~(\ref{kexp}) and Eq.~(\ref{Kexpr2}) that,
\begin{equation}
\delta A_{\mu}= h_{\mu\rho}\delta \tau^{\rho}+ \partial_{\mu}\chi\label{A}
\end{equation}
where $\partial_{\mu}\chi$ represents the U(1) transformation of $A_{\mu}$. Using the expression of $\delta A_{\mu}$ from Eq.~(\ref{A}) the action (\ref{gac2}) simplifies to,
\begin{equation}
\delta S=\int\sqrt{h}d^4x [(J^{\mu}h_{\mu\rho}+R_{\rho})\delta \tau^{\rho}-(\nabla_{\rho}J^{\rho})\chi]
\label{gauge}
\end{equation}
For arbitrary $\chi$, $\delta \tau^{\rho}=0$ gives, 
\begin{equation}
\nabla_{\rho}J^{\rho}=0
\label{nccur}
\end{equation}
This is the equation for the conserved (matter) current in the theory.
For arbitrary $\delta \tau^{\rho}$ and $\chi=0$ we have from Eq.~(\ref{gauge}),
\begin{equation}
R_{\mu}= - J^{\rho}h_{\mu\rho}
\label{spatcurr}
\end{equation}
This is the well known relation between the momentum and particle number currents in non-relativistic theories. 

Considering the variation of the action under diffeomorphisms one has,
\begin{align}
0=\delta S=\int\sqrt{h}d^4x  [-\frac{1}{2}P_{\mu\nu}{\pounds}_{\xi} h^{\mu\nu}+Q^{\mu} {\pounds}_{\xi}\tau_{\mu}+J^{\mu}{\pounds}_{\xi} A_{\mu}+R_{\mu}{\pounds}_{\xi} \tau^{\mu}]\label{genlievar}
\end{align}
where $\pounds_{\xi}$ is the Lie derivative along some arbitrary vector field $\xi^{\mu}$. Eq.~(\ref{genlievar}) can be expressed as,
\begin{align}
0=\delta S=\int \sqrt{h}d^4x~ \xi^{\nu}[\nabla_{\mu}(-T^{\mu}{}_{\nu})+2J^{\mu}\nabla_{[\nu}A_{\mu]}+R_{\mu}\nabla_{\nu}\tau^{\mu}]\label{diffvar}
\end{align}
where
\begin{align}
T^{\mu}{}_{\nu}=P_{\nu\rho}h^{\mu\rho}+Q^{\mu}\tau_{\nu}-R_{\nu}\tau^{\mu}\label{nctmix}
\end{align}
Eq.~(\ref{diffvar}) leads to,
\begin{equation}
\nabla_{\mu}(T^{\mu}{}_{\nu})=2J^{\mu}\nabla_{[\nu}A_{\mu]}+R_{\mu}\nabla_{\nu}\tau^{\mu}=
J^{\mu}K_{\nu\mu}+R_{\mu}\nabla_{\nu}\tau^{\mu}
\label{nabT}
\end{equation}
To provide the constitutive relations we will now describe the physical currents of the theory in terms of fluid variables. For ideal fluids this involves the zeroth order derivative expansion. Since $J^{\mu}$ is some mass flow, we can write
\begin{equation}
J_i^{\mu} = \rho_i u^{\mu}
\label{curr}
\end{equation}
where $\rho_i$ represents the conserved charge density. This choice is by no means exhaustive and in a general derivative expansion for dissipative fluids there exist more terms involving the spatial metric. At zeroth order in the derivative expansion we can also write Eq.~(\ref{nabT}) in the following form,
\begin{equation}
\nabla_{\mu}T^{\mu}{}_{\nu}=\rho h_{\nu\gamma}a^{\gamma}
\label{cons}
\end{equation} 
%This relation differs from the usual relation in relativistic fluid systems. The Newton-Cartan background accounts for additional external forces. $a^{\gamma}$ has been defined previously, and was shown to be the only basic fluid variable which differs in going from inertial to general frames. 
%
We can now deduce the form of $T^{\mu}{}_{\nu}$ for ideal fluids. At this order $P_{\mu \nu} \, , Q^{\mu}$ and $R_{\nu}$ in Eq.~(\ref{nctmix}) will not contain any derivatives of $u^{\mu}$.  %Further, since we are dealing with ideal fluids and hence constant $\rho$, it implies that the expansion $\theta$ vanishes. It also follows from Eq.~(\ref{spatcurr}) that $R_{\nu} = 0$. 
Hence $T^{\mu}{}_{\nu}$ has the following general expression for ideal fluids  
\begin{equation}
T^{\mu}{}_{\nu} = \alpha h_{\nu\rho}h^{\mu\rho}+ \beta u^{\mu}\tau_{\nu} + \gamma h_{\nu \alpha}u^{\alpha}u^{\mu}
\label{idstress}
\end{equation}
By performing the following contractions of  $T^{\mu}{}_{\nu}$ with the expression of Eq.~(\ref{idstress}), 
\begin{equation*}
h^{\nu\alpha}h_{\alpha\mu}T^{\mu}{}_{\nu} = \alpha + \gamma u^{\alpha} u^{\beta}h_{\alpha \beta}\, , \quad \, \tau^{\nu}\tau_{\mu}T^{\mu}{}_{\nu} = \beta \, , \quad \, u^{\nu}\tau_{\mu}T^{\mu}{}_{\nu} = \beta + \gamma u^{\alpha} u^{\beta}h_{\alpha \beta} \, ,
\end{equation*}
we see that $Q^{\mu}$ and $R_{\mu}$ can be interpreted as the energy and momentum currents respectively. This leads to the natural identification of $\beta = \epsilon + \frac{1}{2}\rho u^{\alpha} u^{\beta}h_{\alpha \beta} $ as the total energy of the fluid, $\gamma = - \rho$ to provide the momentum current and $\alpha = -P$. With these conventions for $\alpha\, , \beta$ and $\gamma$ we have
\begin{align}
T^{\mu}{}_{\nu}=(P+ \epsilon + \frac{1}{2}\rho u^{\alpha} u^{\beta}h_{\alpha \beta})u^{\mu}\tau_{\nu}-P\delta^{\mu}_{\nu} - \rho  h_{\nu \alpha}u^{\alpha}u^{\mu}
\label{tupp}
\end{align}

%The expression for $T^{\mu}{}_{\nu}$, in general, is given by Eq. (\ref{nctmix}). 

%\begin{equation}
%T^{\mu\nu}=\epsilon u^{\mu}u^{\nu}-h^{\mu\nu}P\label{tupp}
%\end{equation}

The constitutive relation (\ref{tupp}) for an ideal fluid on the Newton-Cartan background %, which expresses the stress tensor in terms of the energy density $\epsilon$, pressure $P$ and velocity $u^{\mu}$, and is the closest analogue of the relativistic stress energy tensor of an ideal fluid
 is in agreement with the result of \cite{Jensen:2014ama}. The stress tensor of scale invariant fluids on the Newton-Cartan background satisfies the z-deformed trace relation \cite{Christensen:2013lma} $z T^{0}{}_{0} + T^{i}{}_{i} = 0$. With the expression of Eq.~(\ref{tupp}) we find that this trace provides the following {\textit{Equation of state}} when $z=2$,
\begin{equation}
2\epsilon=(n-1)P\label{eos}
\end{equation}
Note that this is a fully classical treatment. If quantum fluids were considered then this relation would follow from the `dilatation Ward identity' associated with the Lifshitz symmetry. %\footnote{This quantum relation is also known as `z-deformed trace' \cite{Christensen:2013lma} in the literature.}.

%Note that Eq.~(\ref{tupp}) represents the physical stress tensor and is not valid under Milne boosts.
 %In a Milne covariant formalism, the velocity $u^{\mu}$ does not Milne transform. This is taken to mean that $u^{\mu}$ is physical.
% is an arbitrary spatial shift vector ($k^{\mu}\tau_{\mu}=0$). While here too Eq.~(\ref{vel}) will be satisfied, in order for Eq.~(\ref{fluvar}) to hold we will need to ensure that $k^{\mu}$ is also irrotational.  Further, we need to modify both $h_{\mu \nu}$ and $A_{\mu}$ to maintain the NC structure in the following manner, 
Let us now consider the velocity $u^{\mu}$ to be Milne invariant. We recall that the set of Milne transformations which leave the (symmetric) connection invariant are, 
%The velocity $u^{\mu}$ does not transform under the Milne transformation as $u^{\mu}$ is considered as a physical field. We recall that the set of Milne transformations that leave the (symmetric) connection invariant are,
\begin{align}
\tau^{\mu} & \to \tau^{\mu} + h^{\mu \nu}k_{\nu} \notag\\
h_{\mu \nu} & \to h_{\mu \nu} - 2 \tau_{(\mu} k_{\nu)} + k^{\alpha} k^{\beta}h_{\alpha \beta} \tau_{\mu} \tau_{\nu} \, , \notag\\
A_{\mu} & \to A_{\mu} +  k_{\mu} - \frac{1}{2}k^{\alpha} k^{\beta}h_{\alpha \beta} \tau_{\mu} \, ,
\label{milne}
\end{align}
where $k_{\mu}$ is an arbitrary spatial vector, i.e. $\tau^{\mu}k_{\mu} = 0$. Under the Milne transformations (\ref{milne}), the variation of (\ref{tupp}) is given by
\begin{equation}
\delta T^{\mu}_{\nu} = -\frac{\rho}{2} u^{\mu}\tau_{\nu} h^{\alpha \beta}k_{\alpha} k_{\beta} + \rho u^{\mu}k_{\nu}
\label{Miltupp}
\end{equation}
There exist several ways in which Milne covariance can be assured. One approach involves redefining $T^{\mu}_{\nu}$ such that
\begin{align}
\tilde{T}^{\mu}{}_{\nu} &= T^{\mu}{}_{\nu} - \rho u^{\mu}A_{\nu} \notag\\
&= (P+ \epsilon + \frac{1}{2}\rho u^{\alpha} u^{\beta}h_{\alpha \beta})u^{\mu}\tau_{\nu}-P\delta^{\mu}_{\nu} - \rho u^{\mu} u^{\beta}h_{\beta \nu} - \rho u^{\mu} A_{\nu}
\label{MinvarianT}
\end{align}
The stress tensor of (\ref{MinvarianT}) is invariant under Milne boosts and agrees with the expression of \cite{Hartong:2016nyx}, where it was derived following the null reduction of a relativistic ideal fluid. A more systematic approach to ensure Milne invariance of all fluid relations to all orders involves the consideration of a Milne covariant formalism. This procedure was first described in \cite{Jensen:2014ama}. Given a Milne invariant velocity $u^{\mu}$, we define $u_{\mu} = h_{\mu \nu}u^{\nu}$ and $u^2 = u_{\mu}u^{\mu}$. We can now replace the Milne variant fields of the Newton-Cartan structure $(h_{\mu \nu}\, , \tau^{\mu} \, , A_{\mu})$ with the new Milne invariant variables $(\tilde{h}_{\mu \nu}\,, u^{\mu}\,,\tilde{A}_{\mu})$, where 
\begin{align}
%u_{\mu} &= h_{\mu \nu}u^{\nu} \notag\\
\tilde{h}_{\mu \nu} &= h_{\mu \nu} - u_{\mu}\tau_{\nu}- u_{\nu}\tau_{\mu} + u^2\tau_{\mu}\tau_{\nu} \notag\\
\tilde{A}_{\mu} &= A_{\mu} + u_{\mu} - \frac{1}{2}\tau_{\mu}u^2 
\label{Minv2}
\end{align}
In this way, beginning with any theory on the Newton-Cartan background, we can transform the gauge variables of the Newton-Cartan structure into the new Milne invariant variables. This is particularly important in the case of the Newton-Cartan background with torsion, since the connection in that case is not simultaneously $U(1)$ and Milne invariant. We will continue to work with the original set of variables of the Newton-Cartan structure as they will allow for a clear relation to the scale covariant Newton-Cartan background to be considered next. %In the resulting equations, we can always transform to the Milne invariant expressions using the transformations just described.
%These are nothing but the familiar Milne transformations for the NC background \cite{Jensen:2014aia}. With these transformations and assumptions, we now find that Eq.~(\ref{cons}) will be satisfied in the new frame provided Eq.~(\ref{tupp}) is modified to,
%\begin{equation}
%T^{\mu}{}_{\nu}=(P+ \epsilon + \frac{1}{2}\rho u^{\alpha} u^{\beta}h_{\alpha \beta})u^{\mu}\tau_{\nu}-P\delta^{\mu}_{\nu} - \rho u^{\mu} u^{\beta}h_{\beta \nu} - \rho u^{\mu} k_{\nu}
%\end{equation}
%This expression for the stress tensor agrees with \cite{Hartong:2016nyx}, where the expression was derived following the null reduction of a relativistic ideal fluid.
 
 Another conservation equation we will be interested in involves the local entropy current. It follows from the second law of thermodynamics as a derived notion. The requirement that entropy should be non-decreasing during hydrodynamic evolution can be
expressed in a covariant way in terms of an entropy current whose divergence is non-negative.
\begin{equation}
\nabla_{\mu}J^{\mu}_S\geq 0\label{enc}
\end{equation}
In Eq.~(\ref{enc}) the equality holds for ideal fluids.
The entropy current $J_S^{\mu}$ can be expressed as,
\begin{equation}
J_{S}^{\mu} = s' u^{\mu}
\end{equation}
where `$s'$' is the entropy density of the fluid. 
\subsection{Fluids on the scale covariant Newton-Cartan background} \label{fluid} 
In this subsection, we first introduce a manifestly Weyl-covariant formalism suited to the study of non-relativistic scale invariant fluids. We assume that our system comprises of tensors $\tilde{Q}^{\alpha...}_{\beta...}$ which obey $\tilde{Q}^{\alpha...}_{\beta...}=e^{ws}Q^{\alpha...}_{\beta...}$, where $w$ is the scaling weight under scale transformations. Correspondingly, we also have the covariant derivative operator $\widetilde{\nabla}$ of the scale covariant Newton-Cartan background which satisfies
\begin{equation}
\widetilde{\nabla}_{\mu} V^{\lambda}_{\rho}=\nabla_{\mu}V^{\lambda}_{\rho}+(b_{\mu}\delta_{\nu}^{\lambda}
+b_{\nu}\delta_{\mu}^{\lambda}-b_{\sigma}h^{\lambda\sigma}h_{\nu\mu})V^{\nu}_{\rho} - (b_{\mu}\delta_{\rho}^{\nu}
+b_{\rho}\delta_{\mu}^{\nu}-b_{\sigma}h^{\nu\sigma}h_{\rho\mu})V^{\lambda}_{\nu}\label{nabv}
\end{equation}
where $\nabla_{\mu}$ is the usual Newton-Cartan covariant derivative and $b_{\mu}$ is the dilatation field. The fluid velocity on the scale covariant Newton-Cartan background transforms as $\tilde{u}^{\mu} = e^{-zs}u^{\mu}$, where $z$ is the dynamical exponent. Given our consideration of the Newton-Cartan background and our interest in the Schr\"odinger field in particular, we will consider the case where $z=2$. However, we will also indicate the results which will follow for general $z$ for many of the subsequent equations. Our analysis will be carried out in $d$ spatial dimensions.
   
Using the above definitions, we can now write the general expression for $\widetilde{\nabla}_{\mu}\tilde{u}^{\nu}$
\begin{equation}
\widetilde{\nabla}_{\mu}\tilde{u}^{\nu} = e^{-zs} \left[(1-z)b_{\mu}u^{\nu} + \nabla_{\mu}u^{\nu} + \left(b_{\alpha}\delta^{\nu}_{\mu} - b_{\sigma}h^{\sigma \nu}h_{\mu \alpha} \right)u^{\alpha}\right]
\label{scalvar}
\end{equation}

With \ref{scalvar} and \ref{fluvar} we find that the expansion, acceleration, shear and vorticity have the following transformations,
\begin{align}
\tilde{\theta}&=\widetilde{\nabla}_{\mu}\tilde{u}^{\mu}=e^{-zs} \left[(d+2-z)b_{\mu}u^{\mu} + \theta\right]\notag\\
\tilde{a^{\nu}}&=\tilde{u}^{\mu}\widetilde{\nabla}_{\mu}\tilde{u}^{\nu}=e^{-2zs} \left[(2-z)b_{\mu}u^{\mu}u^{\nu} + a^{\nu} - b_{\sigma}h^{\sigma \nu}u^2 \right]\notag\\
\widetilde{\sigma}^{\mu\nu}&=e^{-(2+z)s}\left[\sigma^{\mu\nu} + (1-z)b_{\lambda}h^{\lambda (\mu}u^{\nu)} + b_{\lambda}u^{\lambda}h^{\mu \nu} \right] \notag\\
\widetilde{\omega}^{\mu\nu}&=e^{-(2+z)s}\left[\omega^{\mu\nu} + (1-z)b_{\lambda}h^{\lambda [\mu}u^{\nu]} \right]
\label{scalevar}
\end{align}
where $\theta,\, a^{\nu},\, \sigma^{\mu \nu}$ and $\omega^{\mu\nu}$ are defined in \ref{fluvar}.
 
The above set of equations motivate the introduction of a Weyl covariant derivative `${\cal{D}}$' such that for the tensor $\tilde{Q}^{\alpha...}_{\beta...}$ described above, the derivative will act on it as,
\begin{equation}
{\mathcal{D}}\tilde{Q}^{\alpha...}_{\beta...}=e^{ws}{\mathcal{D}}Q^{\alpha...}_{\beta...}
\end{equation}
This leads to the following relation between $\mathcal{D}$ and $\widetilde{\nabla}$ 
\begin{equation}
{\mathcal{D}}_{\mu}=\widetilde{\nabla}_{\mu}- wb_{\mu}\label{wcov}
\end{equation}
%where `$w$' is the conformal weight of the quantity.

Note that the above covariant derivative is metric compatible.
\begin{equation}
{\mathcal{D}}_{\mu}h^{\mu\nu}=0,\, {\mathcal{D}}_{\mu}\tau_{\mu}=0
\end{equation}
For relativistic scale invariant ideal fluids, the Weyl covariant acceleration ($u^{\mu}{\cal{D}}_{\mu}u^{\alpha}$) and expansion (${\cal{D}}_{\mu}u^{\mu}$) are assumed to vanish, leading to an expression for `$b_{\mu}$' in terms of the acceleration and expansion. We can identify a similar relation for the $z=2$ non-relativistic scale invariant fluids using the first two equations of (\ref{scalevar}). The requirements that $u^{\mu}{\cal{D}}_{\mu}u^{\alpha} = 0$ and ${\cal{D}}_{\mu}u^{\mu} = 0$ when $z=2$ can easily be shown to lead to the following relation
\begin{equation}
b_{\mu} = -\frac{\theta}{d} \tau_{\mu} + \frac{a_{\mu}}{u^2}
\label{bz2}
\end{equation}
% We will first treat the action of the conformally invariant derivative `$\mathcal{D}$' on the rescaled tensors as they are before imposing any conditions. Given that the fluid velocity satisfies \ref{vel}, we find the following expression for the acceleration on the rescaled NC background
%\begin{equation}
%u^{\mu}{\cal{D}}_{\mu}u^{\nu} = u^{\mu}\nabla_{\mu}u^{\nu}= a^{\nu}
%\label{acc}
%\end{equation}
%which follows from the fact that the scaling dimension of $u^{\mu}$ is 2.
%We thus see that $u^{\mu}{\cal{D}}_{\mu}u^{\nu} = 0$ when there is no acceleration. Further, the requirement of $\mathcal{D}_{\mu} u^{\mu} = 0$ directly leads to the following condition 
%\begin{equation}
%b_{\mu} u^{\mu} = -\frac{\theta}{n-1}
%\label{shear}
%\end{equation}
%As can be seen from Eq.~(\ref{bz2}) the Weyl covariant derivative is useful in casting the variables and equations of non-relativistic fluid mechanics in a manifestly conformal language. 
These derivatives also define a curvature tensor through their commutator,
\begin{equation}
[{\cal{D}}_{\mu},{\cal{D}}_{\nu}]V^{\lambda}=\tilde{R}^{\lambda}_{\phantom{\lambda} \sigma \mu\nu}V^{\sigma} - \omega F_{\mu\nu}V^{\lambda}
\label{fieldstrength}
\end{equation}
where $F_{\mu\nu}=\tilde{\nabla}_{\mu}b_{\nu}-\tilde{\nabla}_{\nu}b_{\mu}$, and $\tilde{R}^{\lambda}_{\phantom{\lambda}\sigma \mu\nu}$ is as given in Eq.~(\ref{WRiem}). Note that should the usual symmetries of the Riemann tensor be assumed in (\ref{WRiem}), the field strength for the dilatation field $b_{\mu}$ would necessarily vanish. %This is in accordance with the present subsection where the usual symmetries follow through our choice of $b_{\mu} = \partial_{\mu}s$. 
While inconsequential for the case of ideal fluids, $F_{\mu \nu}$ does affect the derivative expansion and dissipative terms which result in non-ideal relativistic fluids \cite{Loganayagam:2008is}. 
%But these equations are not manifestly Weyl covariant as the action of covariant derivative on $T^{\mu\nu}$ has been modified as follows, 

Let us now use this derivative to describe the conservation equations of ideal fluids on the scale covariant Newton-Cartan background. For the stres tensor we consider Eq.~(\ref{cons}) and find that
%The guiding principle will be that the action of the conformally invariant derivative on the rescaled currents of the theory is that same as that of the NC covariant derivative on these currents, whose conservation equations are known. The action of the covariant derivative on the stress tensor can be derived using \ref{nabv}. This leads to the result that,
\begin{equation}
{\cal{D}}_{\mu}T^{\mu}_{\nu} = \rho a_{\nu}
\label{wten}
\end{equation}
provided $T^{\mu}_{\nu}$ has weight `$d+z$' and satisfies $z T^{0}{}_{0} + T^{i}{}_{i} = 0$. It thus follows from Eq.~(\ref{tupp}) that the scaling weights of `$P$' and `$\epsilon$' are both `$d+z$', while the weight for `$\rho$' is `$d+2-z$'. The acceleration `$a_{\nu}$' has the same weight as $u^2$ which is `$2z-2$', with which `$\rho a_{\nu}$' has the expected weight of `$d+z$'. These weights represent the familiar results for Lifshitz fluids.

Likewise for any current $J_i^{\mu} = \rho_i u^{\mu}$ we find that
%\begin{equation}
%\widetilde{\nabla}_{\mu}J^{\mu} = (b_{\mu}\delta_{\nu}^{\mu}
%+b_{\nu}\delta_{\mu}^{\mu}-b_{\sigma}h^{\mu\sigma}h_{\nu\mu})J^{\nu}= (n+1) b_{\mu}J^{\mu} 
%\label{cur}
%\end{equation}
%If the conformal weight of all conserved currents $J_i^{\mu}$ of the theory are $(n+1)$, we then have 
\begin{equation}
\mathcal{D}_{\mu}J^{\mu} = 0
\label{wcur}
\end{equation}
when $J_i^{\mu}$ has the scaling weight of `$(d+2)$'. Thus all current densities $\rho_i$ have the weight `$d+2-z$', including the entropy current relevant to the thermodynamics of the fluid. 
%This result differs from the relativistic case where the weight is always required to be equal to the dimension of the spacetime, i.e. `$n$'. It however does imply that the weight of the density $\rho_i$ be $(n-1)$ just as in the relativistic case, following \ref{curr} and that the weight of $u^{\mu}$ is $2$. For arbitrary anisotropic scaling `$z$', \ref{wcur} will be satisfied if the conformal weight of the current $J^{\mu}_i$ is $(n+z-1)$.

%Apart from the conventional mass flow, we will now consider 
%We now consider the conformal incompressible fluid as a thermodynamic system and 
We assume that our fluid is in local thermodynamic equilibrium in the neighbourhood of any point of spacetime. Let us denote the entropy current density as `$s'$', the temperature as `$T$' and the chemical potentials as `$\mu^i$'. 
%Then from the relation $P + \epsilon = T s' + \mu^i \rho_i$ which should have the uniform weight of $d+z$ it follows that the weight of $T$ is $z$ while that of $\mu^i$ is $2z-2$. 
%In strict analogy to the mass current, there exists a local `entropy current' $J^{\mu}_s$ of the fluid which also has a conformal weight equal to the $(n+1)$. 
%In addition to these equations, we can define an inequality for the entropy current
%\begin{equation}
%\tilde{\nabla}_{\mu}J_S^{\mu} \ge 0
%\label{seclaw}
%\end{equation}
%which follows from the second law of thermodynamics. 
The first law of thermodynamics for this system can be written in terms of the Weyl covariant derivative as,
\begin{equation}
Tu^{\lambda}{\cal{D}}_{\lambda}s'= \frac{(n-1)}{2}u^{\lambda}{\cal{D}}_{\lambda}P-\mu^i u^{\lambda}{\cal{D}}_{\lambda}{\rho}_i
\label{therm}
\end{equation}
It can be noted that the weight of `$T$' is `$z$' while that of `$\mu^i$' be `$2z-2$'. For ideal fluids, from Eq.~(\ref{wten}) and Eq.~(\ref{wcur}) it follows that $u^{\nu}{\cal{D}}_{\mu}T^{\mu}_{\nu} = 0$ and $\mu^i u^{\lambda}{\cal{D}}_{\lambda}{\rho}_i = 0$ respectively. This establishes that $u^{\lambda}{\cal{D}}_{\lambda}P = 0$ on substituting (\ref{tupp}). Thus the entropy density for an ideal fluid on the scale covariant Newton-Cartan background satisfies the following relation
\begin{equation}
T u^{\alpha} {\mathcal{D}}_{\alpha} s'=0
\end{equation}
This implies that the ideal fluids on the scale covariant Newton-Cartan background satisfies the local second law of thermodynamics i.e. the motions of the fluid conserve the entropy of the system and no heat flows in or out of the fluid during its motion. 

Having considered scale invariant ideal fluids in this subsection, we already noted some key differences with the relations that result from relativistic backgrounds. It will be essential to further consider the description of fluids at higher orders in the derivative expansion. This can be carried out using the derivative provided in Eq.~(\ref{wcov}) and the field strength constructed from it in Eq.~(\ref{fieldstrength}), along with the Riemann tensor relations (\ref{WRiem}) - (\ref{tildeRiccs}). In consideing the derivative expansion, we must allow for $\nabla_{[\mu}b_{\nu]} \neq 0$ in general, which will further require us to consider the scale covariant Newton-Cartan background in the presence of torsion. As such, we leave the investigation of this topic to future work.
 \section{Contributions of scale symmetry to the Hall Effect} \label{QHENC}
 %%%%%%%%%%%%%%%%%%%%%%%%%%%%%%%%%%%%%%%%%%

%In this section we will try to motivate the calculation of Hall viscosity for incompressible fluid. As mentioned in the previous section the scale symmetry is an important symmetry when one considers incompressible fluid. For this discussion we will consider the Composite Boson theory,\cite{Zhang1989,Zhang1992} for a Fractional Quantum Hall state. %with filling fraction $\nu=1/(2p+1)$. 
%The corresponding action on curved space will follow from our localization procedure as,

In this section, we will be interested in the consequences of non-relativistic anisotropic scale symmetry in describing Hall fluids. We will follow the procedure described in \cite{Cho:2014vfl} where the Hall viscosity and the Wen-Zee term are derived using an effective hydrodynamic theory. The Hall viscosity results from the Berry phase term in the effective action \cite{Cho:2014vfl}. More specifically, it is the response to spatial stress in the corresponding term of the stress energy tensor. The effective field theory consists of the Schr\"odinger field minimally coupled to a background electromagnetic field $\mathcal{A}_{\mu}$, and a ``dynamical" statistical field $a_{\mu}$. The inclusion of the Chern Simons term involving the field $a_{\mu}$ follows from the need to study perturbations about a mean field of a strongly coupled anyonic system. The field $a_{\mu}$ in effect fixes the statistics of the system to be either bosonic or fermionic and enables the study of responses to the system. After the perturbation has been taken into account, one can then integrate out this field to have the effective field theory description of the Hall fluid. In this context, the field $\Phi$ represents either a composite boson or a composite fermion. Since we are interested in the consequences of curved backgrounds on the system, we will consider the field $\Phi$ as a composite boson. We can then express the Chern Simons Landau Ginzburg (CSLG) effective action of the Quantum Hall system \cite{Zhang:1992eu} in the following way,   
\begin{align}
S =\int dt d^2x  \sqrt{h} &\bigg[ \frac{i}{2} \tau^{\mu} \left(\Phi(x) D_\mu \Phi(x)^{*}  - \Phi^{*}(x)  D_\mu \Phi(x) \right)- \frac{1}{2m}h^{\mu\nu} (D_{\mu}\Phi (x))^{*}(D_{\nu}\Phi (x) )\notag\\& -V(\Phi^*\Phi)+ \frac{\varepsilon^{\mu\nu\lambda}}{8\pi g} a_{\mu}\nabla_{\nu}a_{\lambda} \bigg]
\label{cbaction}
\end{align}
where  $\varepsilon^{\mu \nu \lambda}$ is the Levi Civita tensor, and the covariant derivative on the curved background `$D_{\mu}$' is,
\begin{align} 
D_{\mu} &= \partial_{\mu} + ie\mathcal{A}_{\mu}+ia_{\mu}+ ig\omega_{\mu}+ ig'b_{\mu} \notag\\
& = \partial_{\mu} + i \alpha_{\mu} + ia_{\mu} \, ,
\label{cbcov}
\end{align}
In (\ref{cbcov}) `$\mathcal{A}_{\mu}$' is the external electromagnetic field, `$a_{\mu}$' is the statistical gauge field, `$\omega_{\mu}$' is the field introduced during localization of te Galilean transformations \cite{Banerjee:2014pya} and `$b_{\mu}$' is the dilatation field introduced in Section \ref{SNC}. Following (\ref{gaugefields}) the field $\omega_{\mu}$ can be decomposed as,
\begin{equation}
\omega_{\mu}=\bar{\omega}_{\mu}+mA_{\mu}\label{barbr}
\end{equation} 
where $\bar{\omega}_{\mu}$ is the usual $SO(2)$ connection and the second term in (\ref{barbr}) is related to the mass generating gauge field on Newton-Cartan background. Since we will integrate out the statistical field $a_{\mu}$ before our final result, it will be useful to write the covariant derivative as in the second equality of (\ref{cbcov}). The hydrodynamic version of (\ref{cbaction}) is derived by expressing the complex field $\Phi$ in polar variables \cite{Cho:2014vfl, Zhang:1988wy, Stone}, 
\begin{equation}
\label{a}
\Phi = \sqrt{\rho} e^{i \theta}
\end{equation}
where $\rho=\Phi^{*}\Phi$ is the matter density. The transformation (\ref{a}) leads to the following action,
\begin{align}
S = \int dt d^2x \sqrt{h} [ \rho\tau^{\mu} &\left(\partial_{\mu}\theta +{\alpha}_{\mu}+a_{\mu}\right)-\frac{\rho}{2m}h^{\mu\nu}\left(\partial_{\mu}\theta + {\alpha}_{\mu}+a_{\mu}\right)\left(\partial_{\nu}\theta + {\alpha}_{\nu}+a_{\nu}\right)\notag\\& -\frac{1}{8m\rho}h^{\mu\nu}\partial_{\mu}\rho\partial_{\nu}\rho-V(\rho)+ \frac{\varepsilon^{\mu\nu\lambda}}{8\pi g} a_{\mu}\nabla_{\nu}a_{\lambda}]
\label{calscintaction} 
\end{align}
%where we have denoted $\alpha_{\mu}$ as, $i\alpha_{\mu}=ieA_{\mu}+ iB_{\mu}+ iC_{\mu}$.
The response of the FQH state follows from the variations of the fields 
\begin{align}
\rho &\rightarrow \bar{\rho}+\delta \rho\notag\\\mathcal{A}_{\mu} &\rightarrow \bar{\mathcal{A}}_{\mu}+\delta \mathcal{A}_{\mu}\notag\\ a_{\mu} &\rightarrow \bar{a}_{\mu}+\delta a_{\mu}\label{pertur}
\end{align}
where the barred values represent the mean field values. The FQH state of the electron corresponds to the superfluid state of the boson $\Phi$, where $\bar {\mathcal{A}}_{\mu}$ is completely cancelled by ${\bar a}_{\mu}$ \cite{Cho:2014vfl}. Further, for the Hall fluid the average density, ${\bar \rho}$, is related to the fields $\bar{\mathcal{A}}_{\mu}$ as,    
%In the composite boson theory, the FQH state of the electron corresponds to the superfluid state of the boson $\Phi$. In the superfluid phase, the average ${\bar A}_{\mu}$ of the electromagnetic gauge field $A_{\mu} = {\bar A}_{\mu} + \delta A_{\mu}$ is completely cancelled by the average ${\bar a}_{\mu}$ of the statistical gauge field $a_{\mu} = {\bar a}_{\mu} + \delta a_{\mu}$, {\it i.e.,} ${\bar A}_{i} + {\bar a}_{i}= 0, i =x, y$. 
%On the other hand, the average density of the boson is locked with the average magnetic field due to the quantum Hall effect. 
\begin{align}
{\bar \rho} = \frac{1}{4\pi g} \varepsilon^{0 i j} \nabla_{i}\bar {\mathcal{A}}_{j} (=\frac{1}{4\pi g}B)= -~\frac{1}{4 \pi g} \varepsilon^{0 i j} \nabla_{i}{\bar a}_{j} 
\label{QHE}
\end{align}
where the filling fraction in Eq.~(\ref{QHE}) is written in terms of the intrinsic orbital spin `$g$' through the relation $\nu = \frac{1}{2g}$ and `$B$' is the magnetic field. With these considerations at hand, the effective action (\ref{calscintaction}) provides the following expansion up to quadratic order in variations and derivatives.
%modifies to the following one, where we will retain terms that are at most quadratic in variations and derivatives.  
%We will study the linear response of the following fields,
%where $\bar{\rho}, A, a$ are constant. Therefore we can write down the low-energy Lagrangian for the fluid in terms of $\delta \rho$ and $\theta$ upto quadratic order in the derivatives and the variation in the action,  
\begin{align}
{\mathcal L} = \sqrt{h}&\bigg[ \tau^{\mu}(\partial_{\mu} \theta + \delta \alpha_{\mu}){\bar \rho}   + \tau^{\mu}(\partial_{\mu} \theta + \delta \alpha_{\mu}+\delta a_{\mu})\delta \rho \nonumber\\
&\quad - \frac{{\bar \rho}h^{\mu\nu}}{2m} (\partial_{\mu} \theta + \delta \alpha_{\mu} +\delta a_{\mu}) (\partial_{\nu} \theta + \delta \alpha_{\nu}+\delta a_{\nu})  + \frac{\varepsilon^{\mu\nu\lambda}}{8\pi g} \delta a_{\mu}\nabla_{\nu} \delta a_{\lambda}-V(\bar{\rho}) \bigg]\label{perac}
\end{align}
% Here, the first term $\sim \sqrt{h} {\bar \rho} \partial_{t}\theta$ in the right hand side can be gauged away. Introducing a Hubbard-Stratonovich field $j^{\mu}$ to rewrite the kinetic term of the composite boson the action (\ref{perac}) is modified as,
We can now introduce a field $j^{\mu}$ through a Hubbard-Stratonovich transformation on the kinetic term of the action in Eq.~(\ref{perac}) to rewrite the action as,
\begin{align}
{\mathcal L} = \sqrt{h}&\bigg[  \tau^{\mu}(\partial_{\mu} \theta 
+ \delta \alpha_{\mu}){\bar \rho} + \tau^{\mu}(\partial_{\mu} \theta + \delta \alpha_{\mu}+\delta a_{\mu})\delta \rho - (\partial_{\mu} \theta + \delta \alpha_{\mu}+\delta a_{\mu})h^{\mu\nu}j_{\nu} \nonumber\\ 
&\qquad + \frac{m}{2{\bar \rho}} j_{\mu} h^{\mu\nu}j_{\nu} -V(\bar{\rho}) + \frac{\varepsilon^{\mu\nu\lambda}}{8\pi g} \delta a_{\mu}\nabla_{\nu} \delta a_{\lambda}\bigg],
\label{CB:duality}
\end{align}
%where $J^{\mu} = (\delta \rho, -J^{i})$ represents the conserved boson current.
In the absence of the vortex excitations, we can integrate out the phase variable $\theta$ in (\ref{CB:duality}) to find the following conservation equation,
\begin{equation}
\partial_{\mu} (\sqrt{h}J^{\mu}) = \sqrt{h} \nabla_{\mu}J^{\mu} =0
\label{jcon}
\end{equation}
where we have defined $J^{\mu} = \delta \rho \tau^{\mu} - j_{\nu}h^{\nu \mu}$. Given Eq.~(\ref{jcon}) holds, we can further express it as,
\begin{equation}
J^{\mu} = \varepsilon^{\mu\nu\lambda} \frac{1}{2\pi} \nabla_{\nu}f_{\lambda} \, , 
\end{equation}
where $f_{\lambda}$ are the new hydrodynamic gauge variables. Clearly, $J^{\mu}$ remains invariant under $U(1)$ transformations of the field $f_{\lambda}$. By substituting this expression for $J^{\mu}$ back in \ref{CB:duality}, we find,
% and have expressed it in terms of the  hydrodynamic (gauge) field $b_{\mu}$ due to its conservation. By plugging this back to the Lagrangian Eq. \eqref{CB:duality}, we find  
\begin{align}
{\mathcal L} =& \sqrt{h}\bigg[  {\bar \rho}\tau^{\mu} \delta \alpha_{\mu} + \varepsilon^{\mu\nu\lambda} \frac{1}{2\pi} \nabla_{\nu}f_{\lambda}(\delta \alpha_{\mu}+\delta a_{\mu}) +\frac{m}{2{\bar \rho}} j_{\mu} h^{\mu\nu}j_{\nu} + \frac{\varepsilon^{\mu\nu\lambda}}{8\pi g} \delta a_{\mu}\nabla_{\nu} \delta a_{\lambda}-V(\bar{\rho}) \bigg]  \,
\label{CB:dual}
\end{align}
%\begin{align}
%&{\mathcal L}  =\sqrt{g} \bigg[ {\bar \rho} \delta \alpha_{t} + \frac{1}{2\pi} \varepsilon^{\mu\nu\lambda} (\delta \alpha_{\mu} + %\delta a_{\mu})\nabla_{\nu}b_{\lambda} \nonumber\\ 
%&\quad +\frac{\varepsilon^{\mu\nu\lambda}}{4\pi (2p+1)} \delta a_{\mu}\nabla_{\nu} \delta a_{\lambda} - \frac{1}{2{\bar \rho}} e_{i} g^{ij}e_{j} \bigg]. 
%\end{align}
%Here $e_{i} = \frac{1}{2\pi} \varepsilon^{i\sigma\lambda}\nabla_{\sigma}b_{\lambda}, i=x,y$ is the electric field of $b_{\mu}$. 

Integrating out $\delta a_{\mu}$ and using the expression of $\delta {\alpha_{\mu}}$ from Eq.~(\ref{cbcov}) we obtain this effective theory for the Hall state on the scale invariant Newton-Cartan background upto the leading order in gauge fields,
\begin{align}
{\mathcal L} = & \sqrt{h} \left[(g \tau^{\mu} \omega_{\mu} {\bar \rho} + g'\tau^{\mu} b_{\mu} {\bar \rho})+( \frac{g}{2\pi} \varepsilon^{\mu\nu\lambda} \omega_{\mu} \partial_{\nu}f_{\lambda}+ \frac{g'}{2\pi} \varepsilon^{\mu\nu\lambda} b_{\mu} \partial_{\nu}f_{\lambda}) \right.  \notag\\ & \left. \qquad \qquad +e \tau^{\mu} \delta \mathcal{A}_{\mu}  \bar \rho +  \frac{e}{2\pi} \varepsilon^{\mu\nu\lambda} \delta \mathcal{A}_{\mu} \partial_{\nu}f_{\lambda}-\frac{g}{2\pi} \varepsilon^{\mu\nu\lambda} f_{\mu}\partial_{\nu} f_{\lambda}  +\cdots \right]
\label{FQHE}
\end{align}
The first parenthesis in Eq.~(\ref{FQHE}) represents the Berry phase terms and the terms in the second parenthesis are the Wen-Zee terms. The terms with coefficient `$g$' arise due to the symmetries of the Newton-Cartan background. The terms involving `$g'$' are the contributions due to the scale covariance of the background. Our aim is to study the response of the effective action (\ref{FQHE}) to time dependent variations of the spatial metric. This response receives contributions only from those terms which are quadratic in variations of the spatial metric under the presence of a constant magnetic field ($\bar{\rho} =$ const.). Hence only the Berry phase terms will be relevant to study the contribution to the Hall viscosity through the stress tensor. The Wen-Zee terms will change the flux due to the curved background in a time independent manner.

We will consider the time dependent variations of the spatial metric and its inverse about flat space, which we will label as $\delta h_{\mu \nu}(t)$ and $\delta h^{\mu \nu}(t)$ respectively. In doing so with Eq.~(\ref{FQHE}), we end up with the following contribution which is quadratic in variations,
\begin{equation}
L_2 = \frac{1}{8} g \bar{\rho} \, \epsilon_{a b} \delta^{a \mu} \delta^{b}_{\nu} \left(\delta h_{\mu \rho} \delta \dot{h}^{\rho \nu} \right) + \frac{1}{4} g' \bar{\rho} \delta h_{\mu \rho}\delta \dot{h}^{\mu \rho} + \cdots
\label{eff}
\end{equation}
where the overdot implies the time derivative and `$\cdots$' denotes those terms other than quadratic order, which have been neglected in $L_2$. Using Eq.~(\ref{eff}), we find the following correction to the stress tensor,
\begin{equation}
T^{\mu}_{\nu} = \frac{\eta_H}{2} \left(\frac{1}{2}\epsilon_{a b} \delta^{a \mu} \delta^{b}_{\sigma} \delta h_{\lambda \nu} \dot{\delta h}^{\lambda \sigma} - \frac{1}{2}\epsilon_{a b} \delta^{a \sigma} \delta h_{\sigma \lambda} \delta^{b}_{\nu} \dot{\delta h}^{\lambda \mu} \right) + \frac{\theta_H}{2} \partial_t\left(\delta h^{\mu \sigma} \delta h_{\sigma \nu} \right)
\label{stress2}
\end{equation}
where we have denoted $\frac{g \bar{\rho}}{2} = \eta_H$ and $\frac{g' \bar{\rho}}{2} = \theta_H$. `$\eta_H$' is the Hall viscosity and `$\theta_H$' is the correction to the Hall viscosity which follows from our consideration of the scale covariant Newton-Cartan background. As $\bar{\rho}$ is proportional to the magnetic field $B$ following (\ref{QHE}), both the Hall viscosity and the correction to it are also proportional to the magnetic field. In deriving \ref{stress2} we made use of the fact that $\epsilon_{0 b} = 0$.  In involving the time derivative of the spatial metric variations this additional term rescales the Hall fluid. We note that the spatial metric variations must also be related to corresponding temporal variations of $\tau^{\mu}$, so as to satisfy (\ref{ncproj}). As such, this term may also be viewed as an expansion of the Hall droplet which results in order to preserve the scale covariance of the Newton-Cartan background. 

Scale invariance can be expected to play a larger role in the context of Hall fluids on Newton-Cartan backgrounds. It is known, for instance, that the inclusion of torsion affects the response of Quantum Hall systems \cite{Gromov:2014vla}. Since the scale covariant Newton-Cartan background is known to constrain the expression of the torsion tensor (\ref{torfull}), its inclusion in the above analysis will lead to additional scale dependent responses. We also noted that variations of the temporal vierbein will be necessary in order to ensure that the spatial metric variations preserve the Newton-Cartan structure.
% By Eq. (\ref{tupp}), such temporal variations will lead to their own responses to quadratic order. 
This is a feature of the Newton-Cartan background, as such constraints on metric variations do not arise in the analysis of the Quantum Hall effect on relativistic backgrounds. We leave the detailed analysis of such responses to future work, which should lead to further insight into the term derived in Eq.~(\ref{stress2}).

%%%%%%%%%%%%%%%%%%%%%%%%%%%%%%%%%%%%
\section{Conclusion} \label{Conc}
%%%%%%%%%%%%%%%%%%%%%%%%%%%%%%%%%%%%%%
In this paper the scale transformation was successfully included in the localization scheme for a non-relativistic field theoretic model. As a consequence, we can now consistently couple massive Galilean plus dilatation invariant Schr\"odinger fields to a curved background. The background was identified as the scale covariant Newton-Cartan background. Following this we derived the connection, Riemann tensor, Ricci scalar and Weyl tensor for the scale covariant background. The connection of this background differs from that of the Newton-Cartan one by terms which involve the dilatation field corresponding to anisotropic scale transformations. We also demonstrated that while the Riemann and Ricci tensors scale quite differently than their relativistic counterparts, the definitions of the Weyl and Schouten tensors are of a similar form as in the relativistic case. 

We then provided the description of scale invariant ideal fluids on this background. The weights of the energy, entropy and matter current densities were derived, through which we were further able to demonstrate the first law of thermodynamics. An interesting feature of describing fluids on this background is the possible inclusion of external forces in a geometric framework, which can result from specific choices of the field $A_{\mu}$. While we have considered the case of an ideal fluid, it is clear that we will need to further consider viscous and dissipative scale invariant fluids as well. The entropy current in these cases is non-trivial, and a second order derivative expansion must be carried out. The investigation of these topics requires the relations and derivatives introduced in section \ref{fluid}, as well as the expressions of the curvature tensors provided in Eqs.~(\ref{WRiem}-\ref{tildeRiccs}). 

We further considered the consequences of scale invariance in an effective field theory of Hall fluids. The inclusion of the dilatation field was shown to modify the low energy hydrodynamic effective field theory. By considering spatial metric perturbations, we determined that the corresponding response involved an expansion of the Hall fluid. In addition, we discussed that in considering the scale covariant Newton-Cartan background, it is in general inconsistent to merely introduce spatial metric variations without the consideration of corresponding variations in the temporal metric, so as to preserve the orthogonality relations between the two. This property enables the scale corrections to the effective action to contribute through metric perturbations. The torsion tensor of the scale covariant Newton-Cartan background, through its dependence on the temporal metric and dilatation field, promise additional geometric terms which could be introduced in the low energy effective action. While a detailed investigation of these terms and their phenomelogical consequences are interesting, they lie beyond the scope of the present work.

\appendix
\section{Derivation of the Schouten tensor}\label{app1}
The key property of $C_{\lambda\sigma\mu\nu}$ is that it vanishes when any pair of indices is contracted with either $h^{\mu \nu}$ or $\tau^{\mu} \tau^{\nu}$. By requiring that $C_{\lambda\sigma\mu\nu}$ be trace free, the expression for $S_{\nu \sigma}$ can be derived. For the moment we will assume that the Riemann tensor of the Newton-Cartan background satisfies the conditions in (\ref{ncRSsymm}) and (\ref{trautman}), and that the connection is symmetric. Due to the first equality in Eq.~(\ref{ncRSsymm}), we have 
\begin{equation}
R_{\lambda \sigma \mu \nu} = h_{\lambda \epsilon}R^{\epsilon}_{\phantom{\epsilon} \sigma \mu \nu}\label{rr}
\end{equation}
%But the Riemann tensor provides a non-vanishing result when contracted with both $h^{\mu \nu}$ and $\tau^{\mu} \tau^{\nu}$, depending on the indices being contracted. Requirement of vanishing $C_{\lambda \sigma \mu \nu}$ upon contraction of any two of its indices with either $h^{\mu \nu}$ or $\tau^{\mu} \tau^{\nu}$ results an expression of Schouten tensor.
It further follows from the symmetries that 
\begin{align}
h^{\lambda \mu}R_{\lambda \sigma \mu \nu} &= R_{\sigma \nu}  \label{app2.r1} \\
\tau^{\lambda}\tau^{\mu}R_{\lambda \sigma \mu \nu} &= 0 \label{app2.r2} \\
h^{\sigma \mu}R_{\lambda \sigma \mu \nu} &= 0 \label{app2.r4}
\end{align}
These relations represent the covariant description of the familiar result that the only non-vanishing component of the Newton-Cartan Riemann tensor is $R^{i}_{\phantom{i} 0 j 0}$ \cite{Misner:1974qy}. We can now use these relations to determine $S_{\nu \sigma}$. Contracting Eq.~(\ref{ncweyl}) with $(h^{\lambda \mu}+\tau^{\lambda}\tau^{\mu})$ gives,
\begin{equation}
R_{\sigma \nu} = (n-2)S_{\nu \sigma} + (h^{\lambda \mu} + \tau^{\lambda} \tau^{\mu})S_{\lambda \mu}(h_{\sigma \nu} + \tau_{\sigma} \tau_{\nu})
\label{app2.res1}
\end{equation}
where we have used Eq.~(\ref{app2.r1}) and Eq.~(\ref{app2.r2}). Contracting Eq.~(\ref{app2.res1}) with $h^{\sigma \nu}$ and $\tau^{\sigma} \tau^{\nu}$ respectively we find
\begin{align}
\tau^{\nu} \tau^{\sigma} S_{\nu \sigma} = - S_{\nu \sigma} h^{\nu \sigma} \frac{(2n-3)}{(n-1)} \, , \quad
R = (n-1) \tau^{\nu} \tau^{\sigma} S_{\nu \sigma} + S_{\nu \sigma} h^{\nu \sigma} \label{app2.res3}
\end{align}
From Eq.~(\ref{app2.res3}) we thus find,
\begin{equation}
S_{\nu \sigma} h^{\nu \sigma} = - \frac{R}{2 (n-2)}
\label{app2.res4}
\end{equation}
Substituting $\tau^{\nu} \tau^{\sigma} S_{\nu \sigma}$ and $S_{\nu \sigma} h^{\nu \sigma}$ from Eq.~(\ref{app2.res3}) and Eq.~(\ref{app2.res4}) in (\ref{app2.res1}) gives the following expression for $S_{\nu \sigma}$,
\begin{equation}
S_{\nu \sigma} = \frac{1}{n-2} \left(R_{\sigma \nu} - \frac{1}{2 (n-1)} R (h_{\sigma \nu} + \tau_{\sigma} \tau_{\nu}) \right)\label{ncSch} 
\end{equation}
It is evident from Eq.~(\ref{ncSch}) that $S_{\nu \sigma}$ has a form similar to that of the Schouten tensor in General Relativity.
%\begin{align}
%h_{\mu\nu}\tau^\nu &=  {\Lambda_{\mu}}^a {\Lambda_\nu}^a {\Sigma_0}^\nu\notag\\ &=  {\Lambda_{\mu}}^a\delta_0^a\notag\\ &=0
%\end{align}In a similar manner the following projection relations,Given these relations, it is clear that despite the rescaled Newton-Cartan geometry generating non-metricity, the familiar relations of Newton-Cartan geometry continue to hold.

%The description of the curved background is as yet incomplete, in that the explicit form of the connection involved in the covariant derivatives has not yet been derived. 

%\section{Schouten tensor of the Newton-Cartan background} \label{Schouten}
%\section{Acknowledgment}
%I would like to thanks Prof. Michael Stone, Prof. Rabin Banerjee and Prof. Pradip Mukherjee for useful discussions.

\end{document}